\documentclass[camera,letterpaper,nomarginnotes, 10pt, nonarrowgutter]{jpaper}
\usepackage[table]{xcolor}
\usepackage[T1]{fontenc}
\usepackage{booktabs}
\usepackage[linesnumbered,ruled]{algorithm2e}

\usepackage{listings}
\usepackage{fancyhdr}
\usepackage{datetime}
\usepackage{cite}
\usepackage{amsmath,amssymb,amsfonts}
\usepackage{algorithmic}
\usepackage{graphicx}
\usepackage{textcomp}
\usepackage[table]{xcolor}
\usepackage{tikz}
\usepackage[utf8]{inputenc}
\usepackage[T1]{fontenc}
\usepackage{booktabs} %
\usepackage{setspace}
\usepackage[italic]{mathastext}
\usepackage{array}
\usepackage{titlesec}
\usepackage[normalem]{ulem}
\usepackage{multirow}
\usepackage{multicol}
\usepackage{color}
\usepackage[font={small,bf}]{caption}
\usepackage{float}
\usepackage[font={small,bf}]{subcaption}
\usepackage[linesnumbered,ruled]{algorithm2e}
\usepackage{makecell}
\usepackage{pifont}
\usepackage[]{microtype}

\usepackage{soul} %

\usepackage[colorinlistoftodos,prependcaption,textsize=small]{todonotes}
\usepackage{marginnote} 

\usepackage{clipboard}
\usepackage{balance}
\usepackage[framemethod=tikz]{mdframed}
\usepackage{xargs}
\usepackage[most]{tcolorbox}
\usepackage{cancel}

\usepackage[bookmarks=true,breaklinks=true,colorlinks,linkcolor=black,citecolor=blue,urlcolor=black]{hyperref}

\usepackage[resetlabels]{multibib}

\definecolor{denim}{rgb}{0.08, 0.38, 0.74}
\definecolor{darkolivegreen}{rgb}{0.33, 0.42, 0.18}
\definecolor{dgreen}{rgb}{0.00, 0.75, 0.00}
\definecolor{darkpink}{rgb}{0.88, 0.28, 0.54}
\definecolor{forestgreen}{rgb}{0.0, 0.27, 0.13}
\definecolor{amber}{rgb}{1.0, 0.49, 0.0}
\definecolor{lightyellow}{rgb}{0.980, 0.956, 0.623}
\definecolor{lightblue}{rgb}{0.980, 0.956, 0.623}
\definecolor{darkamber}{rgb}{0.5, 0.19, 0.0}
\definecolor{dkgreen}{rgb}{0,0.6,0}
\definecolor{gray}{rgb}{0.5,0.5,0.5}
\definecolor{mauve}{rgb}{0.58,0,0.82}
\definecolor{lightmauve}{rgb}{0.68,0.4,0.92}
\definecolor{chocolate}{rgb}{0.55, 0.32, 0.09}
\definecolor{dollarbill}{rgb}{0.52,0.73,0.4}
\definecolor{dkdkgreen}{rgb}{0,0.45,0}
\definecolor{gfored}{rgb}{0.580, 0.050, 0.211}
\definecolor{darkwarmgray}{rgb}{0.15, 0.050, 0.05}
\definecolor{ups-truck}{rgb}{0.53, 0.28, 0.21}

\definecolor{bestresult}{HTML}{b3e2cd} %
\definecolor{worseresult}{HTML}{f4c7c3} %

\makeatletter
\g@addto@macro{\normalsize}{%
  \setlength{\abovedisplayskip}{2pt plus 1pt minus 1pt}
  \setlength{\belowdisplayskip}{2pt plus 1pt minus 1pt}
  \setlength{\intextsep}{2pt plus 1pt minus 1pt}
  \setlength{\textfloatsep}{3pt plus 1pt minus 1pt}
  \setlength{\dbltextfloatsep}{3pt plus 1pt minus 1pt}
  \setlength{\skip\footins}{4pt plus 1pt minus 1pt}
}
\def\BibTeX{{\rm B\kern-.05em{\sc i\kern-.025em b}\kern-.08em
    T\kern-.1667em\lower.7ex\hbox{E}\kern-.125emX}}

\def\UrlBreaks{\do\/\do-\/\do.\/\do:}

\expandafter\def\expandafter\UrlBreaks\expandafter{\UrlBreaks
  \do\/\do\.\do\-\do\:}
  
\newcommand{\squishlist}{
 \begin{list}{$\circ$}
  { \setlength{\itemsep}{0pt}
     \setlength{\parsep}{0pt}
     \setlength{\topsep}{0pt}
     \setlength{\partopsep}{0pt}
     \setlength{\leftmargin}{1em}
     \setlength{\labelwidth}{1em}
     \setlength{\labelsep}{0.5em} } }

\newcommand{\squishsublist}{
\begin{list}{$\rightarrow$}
 { \setlength{\itemsep}{0pt}
    \setlength{\parsep}{0pt}
    \setlength{\topsep}{-10em}
    \setlength{\partopsep}{-3pt}
    \setlength{\leftmargin}{1em}
    \setlength{\labelwidth}{1em}
    \setlength{\labelsep}{0.5em} } }

\newcommand{\squishend}{
  \end{list}  }

\newcommand{\head}[1]{\noindent\textbf{#1.}} %

\newcommand{\circled}[1]{{\tikz[baseline=(char.base)]{\node[shape=circle,inner sep=1.3pt,fill=black, text=white] (char) {\small \textbf{#1}};}}}

\usepackage{titlesec}
\titlespacing*{\section}{0pt}{0.3ex}{0.1ex} %
\titlespacing*{\subsection}{0pt}{0.3ex}{0.3ex} %
\titlespacing*{\subsubsection}{0pt}{0.3ex}{0.3ex} %

\newcommand\mech{CRANE\xspace}
\newcommand\ltitle{\mech: Correcting Errors in Raw Nanopore Signals\\Using Hidden Markov Models\xspace}

\newcommand{\release}{\href{https://github.com/STORMgroup/CRANE}{\url{https://github.com/STORMgroup/CRANE}}\xspace}

\newcommand{\citenanopore}{menestrina_ionic_1986,cherf_automated_2012,manrao_reading_2012,laszlo_decoding_2014,deamer_three_2016,kasianowicz_characterization_1996,meller_rapid_2000,stoddart_single-nucleotide_2009,laszlo_detection_2013,schreiber_error_2013,butler_single-molecule_2008,derrington_nanopore_2010,song_structure_1996,walker_pore-forming_1994,wescoe_nanopores_2014,lieberman_processive_2010,bezrukov_dynamics_1996,akeson_microsecond_1999,stoddart_nucleobase_2010,ashkenasy_recognizing_2005,stoddart_multiple_2010,bezrukov_current_1993,zhang_single-molecule_2024}

\newcommand{\citebasecallnanodnn}{cavlak_targetcall_2024,xu_fast-bonito_2021,peresini_nanopore_2021,boza_deepnano_2017,boza_deepnano-blitz_2020,oxford_nanopore_technologies_dorado_2024,oxford_nanopore_technologies_guppy_2017,lv_end--end_2020,singh_rubicon_2024,zhang_nanopore_2020,xu_lokatt_2023,zeng_causalcall_2020,teng_chiron_2018,konishi_halcyon_2021,yeh_msrcall_2022,noordijk_baseless_2023,huang_sacall_2022,miculinic_mincall_2019}

\newcommand{\citesignalanalysis}{currentview2026, graphlearningevent2026, eventdetect_wei_2026, nanolabel2026, eris_rawbench_2025,bao_squigglenet_2021,loose_real-time_2016,zhang_real-time_2021,kovaka_targeted_2021,senanayake_deepselectnet_2023,sam_kovaka_uncalled4_2025,lindegger_rawalign_2023,firtina_rawhash_2023,firtina_rawhash2_2024,firtina_rawsamble_2026,shih_efficient_2023,sadasivan_rapid_2023,dunn_squigglefilter_2021,shivakumar_sigmoni_2024,sadasivan_accelerated_2024,gamaarachchi_gpu_2020,samarasinghe_energy_2021, soysal_mars_2025}

\newcommand{\citehwbasecgpu}{cavlak_targetcall_2024,xu_fast-bonito_2021,boza_deepnano_2017,oxford_nanopore_technologies_bonito_2021,oxford_nanopore_technologies_dorado_2024,oxford_nanopore_technologies_guppy_2017,lv_end--end_2020,singh_rubicon_2024,zhang_nanopore_2020,xu_lokatt_2023,zeng_causalcall_2020,teng_chiron_2018,konishi_halcyon_2021,yeh_msrcall_2022,noordijk_baseless_2023,huang_sacall_2022,sneddon_language-informed_2022,miculinic_mincall_2019}

\def\BibTeX{{\rm B\kern-.05em{\sc i\kern-.025em b}\kern-.08em
    T\kern-.1667em\lower.7ex\hbox{E}\kern-.125emX}}
    
\newcites{supp}{Supplementary References}

\begin{document}

\title{\ltitle}

\newcommand{\affila}[0]{\small {$^1$}}
\newcommand{\affilb}[0]{\small {$^2$}}
\newcommand{\affilc}[0]{\small {$^3$}}
\newcommand{\affild}[0]{\small {$^4$}}
\newcommand{\affile}[0]{\small {$^5$}}
\author{
\vspace{-18pt}\\%
{Simon Ambrozak\affila{}}\quad%
{Ulysse McConnell\affila{}\textsuperscript{,}\affilb{}}\quad%
{Bhargav Srinivasan\affila{}}\quad \\%
{Burak Ozkan\affila{}\textsuperscript{,}\affilc{}}\quad%
{Ernest Zhang\affila{}}\quad%
{Can Firtina\affila{}}\quad%
\vspace{-1pt}\\%
\affila\emph{University of Maryland}%
\quad
\affilb\emph{ETH Zurich}%
\quad
\affilc\emph{Bilkent University}%
}

\maketitle
\pagestyle{plain}
\thispagestyle{plain}
\begin{abstract}
{\noindent \textbf{Abstract:} Nanopore sequencing can read substantially longer sequences of nucleic acid molecules, called reads, than other sequencing methods, which has led to advances in genomic analysis such as the gapless human genome assembly. By analyzing the raw electrical signal reads that nanopore sequencing generates from molecules, existing works can map these reads without translating them into DNA characters (i.e., basecalling), allowing for quick and efficient analysis of sequencing data. However, raw signals often contain errors due to noise and processing errors, which limits the overall accuracy of raw signal analysis.

Our goal in this work is to detect and correct errors in raw signals to improve the accuracy of raw signal analyses. To this end, we propose \mech, a mechanism that trains and utilizes a Hidden Markov Model (HMM) to accurately correct signal errors. Our extensive evaluation on various datasets shows that \mech 1)~consistently improves the overall accuracy of the underlying raw signal analysis tools, 2)~minimizes the burden of optimizing analysis pipelines for newer nanopore technologies, and 3)~does not introduce substantial computational overhead. We conclude that \mech provides an effective mechanism to systematically identify and correct the errors in raw nanopore signals before further analysis, which can enable the development of a new class of error correction mechanisms purely designed for raw nanopore signals.
\\\textbf{Source Code:} \mech is available at \release. We also provide the scripts to fully reproduce our results on our GitHub page. \\
}
\end{abstract}

\section{Introduction} \label{xyz:sec:introduction}
Nanopore sequencing technology has enabled the high-throughput sequencing of very long nucleic acid molecules (e.g., DNA), called \emph{long reads}, often exceeding thousands of bases in length~\cite{\citenanopore}. These long reads are particularly useful for many applications in genomics such as identifying complex and repetitive regions of genomes~\cite{noyes2026longread}, and constructing gapless assemblies~\cite{nurk2022genome}. To sequence these long reads, nanopore sequencing produces series of noisy \emph{electrical signals} based on the ionic current disruptions that nucleic acid molecules generate as each nucleotide passes through a nanometer-scale pore, called a \emph{nanopore}.

Apart from the capabilities to produce \emph{ultra} long reads up to a few million bases, nanopore sequencing provides two unique benefits. First, nanopore sequencing enables stopping the sequencing process of a read (i.e., Read Until~\cite{loose_real-time_2016}) or the entire sequencing run (i.e., Run Until~\cite{payne_readfish_2021}) without fully sequencing it, a technique known as \emph{adaptive sampling}. This can significantly reduce sequencing time and cost by better utilizing sequencing consumables, called flow cells. To decide if a sequencing process should stop early, tools must analyze raw nanopore signals as these signals are generated in \emph{real-time}. Second, the small scale of nanopores allows for portable handheld sequencers which can be used in mobile and resource-constrained environments without access to cloud computing. Such situations might require minimal computational latency for effective adaptive sampling.
With capabilities including ultra long reads, adaptive sampling and portable sequencing, many analysis pipelines use nanopore sequencing for various applications such as telomere-to-telomere gapless genome assembly~\cite{nurk2022genome}, metagenomics~\cite{zhong_metagenomics_2024}, complex structural variant detection~\cite{stancu2017mapping}, and in-the-field analyses such as continuous outbreak tracing~\cite{quickPortable2016}.

To enable many of these applications, raw nanopore electrical signals are mainly analyzed in two ways. First, the most common approach is to translate these raw nanopore signals into human-readable nucleotide sequences with a computational step called \emph{basecalling}. To accurately translate from noisy electrical signals to nucleotide sequences, these basecalling techniques commonly rely on complex machine learning (ML) models~\cite{\citebasecallnanodnn} that usually combine several layers of CNNs~\cite{teng_chiron_2018}, transformers~\cite{huang_sacall_2022}, and decoders such as CRF~\cite{oxford_nanopore_technologies_bonito_2021} and CTC~\cite{teng_chiron_2018}. However, these approaches are usually costly and require resource-intensive devices such as GPUs~\cite{\citehwbasecgpu}. Second, to avoid the significant computational demand of basecalling, several approaches~\cite{\citesignalanalysis} directly analyze raw nanopore signals \emph{without} basecalling them. These raw signal analysis approaches provide substantial benefits in terms of the computational resources they require compared to a pipeline that uses computationally costly basecalling. Reducing the computational overhead is particularly useful for scalability (i.e., how many processing units or CPU threads needed to process the entire flow cell in real-time), latency (i.e., how quickly a real-time decision can be made), and energy reasons (e.g., what is the required power draw from a mobile device).

Although existing raw signal analysis approaches provide substantial benefits in terms of lower computational overhead, they generally exhibit lower accuracy than the analysis pipelines that use basecalling. This is mainly because 1)~basecalling techniques are heavily optimized for very accurate translation from noisy electrical signals and 2)~the signal processing algorithms used in raw signal analysis are prone to making errors, which usually propagates to the later steps in signal analysis and reduces the overall accuracy of these approaches.

A common initial processing step in raw signal analyses aims to identify \emph{segments} in electrical signals, also called \emph{events}, before further processing the signals. Each event usually corresponds to the series of electrical signals measured when a particular sequence of nucleotides of a fixed length \texttt{k} (i.e., k-mer) pass through a nanopore. By identifying events, raw signal analysis techniques can differentiate between signal regions that correspond to each successive k-mer of a nucleic acid molecule for further processing. This is because signal amplitudes within the same event are \emph{expected} to be mutually similar, whereas amplitudes originating from distinct events (i.e. different k-mers) are dissimilar. Existing raw signal analysis tools exploit this property to perform k-mer matching directly in the raw signal space by employing heuristics to mitigate signal noise. For example, the state-of-the-art raw signal mapping tool, RawHash2~\cite{firtina_rawhash2_2024}, uses the average signal value of each event to generate a hash value from several consecutive quantized event values before performing a hash-based seeding. 

Existing \emph{lightweight} segmentation algorithms, such as the rolling t-test in Scrappie~\cite{oxford_nanopore_technologies_scrappie_2019}, aim to find statistically significant changes within a window of signals to identify the borders of events, known as segmentation points. However, these approaches inherently generate a large number of spurious events from a raw signal, often called \emph{oversegmentation} errors~\cite{shivakumar_sigmoni_2024}, and they require optimizing their parameters depending on the nanopore versions (i.e., chemistries such as R9.4 and R10.4.1)~\cite{bakic_campolina_2026}, which heavily impacts the accuracy of downstream analysis (discussed in Section~\ref{cern:subsec:results}). 
Due to the significant impact that the oversegmentation errors cause in raw signal analyses, a recent work, Campolina~\cite{bakic_campolina_2026}, proposes a deep learning-based design for accurately identifying segmented regions while reducing the oversegmentation issues. Such deep learning-based works have potential in both 1)~performing more accurate segmentation and 2)~subsequently improving the downstream raw signal analysis. However, these deep learning approaches are mainly practical when using GPUs, as their CPU executions are substantially slower than lightweight statistical segmentation algorithms, and they can still introduce errors~\cite{bakic_campolina_2026}.

To naively and quickly correct these oversegmentation errors, the most common approach used in several raw signal works~\cite{zhang_real-time_2021, firtina_rawhash_2023, firtina_rawhash2_2024, shivakumar_sigmoni_2024, firtina_rawsamble_2026} is to perform homopolymer compression (HPC) to quickly identify consecutive events with similar signal amplitudes and \emph{compress} them into one (e.g., by picking the first event in a homopolymer run of such events), similar to error correction mechanisms in assembly~\cite{miller_aggressive_2008} and in read mapping tools~\cite{li_minimap2_2018} for basecalled sequences. Although HPC significantly improves the overall raw signal analysis by reducing \emph{some} of the oversegmentation errors, it can remove error-free and informative events, and it is limited to correcting only oversegmentation errors, leaving other types of errors such as noise unhandled. To our knowledge, there is no error correction mechanism specifically designed to correct various types of errors in segmented raw signals.

Our goal is to substantially improve the accuracy of raw signal analysis approaches that rely on segmentation algorithms by 1)~correcting the errors that these algorithms make and 2)~reducing their dependency on optimized parameter setting for different nanopore chemistries. To this end, we propose \mech, the \emph{first} mechanism that corrects raw signal events by learning from error-free nanopore event sequences. To effectively model event sequences, \mech uses probabilistic graph structures, \emph{Hidden Markov Models} (HMMs), which are trained and utilized in three key steps. First, to learn from error-free event sequences, we train an HMM to model error-free synthetic data. Second, to adapt \mech for use in different raw signal analysis pipelines, it uses three parameters: HMM size, $P(\text{stay})$, and $P(\text{skip})$, which correspond to the number of HMM states, oversegmentation rates and model uncertainty, respectively. These parameters are then optimized for a specific pipeline using a parameter search step guided by evaluating the pipeline across different parameter combinations. Third, \mech uses a modified decoding algorithm that finds the most likely path through the trained HMM, called the \emph{Viterbi algorithm}, to identify and correct errors in real-world nanopore event sequences.

Our extensive evaluations on various segmentation algorithms using \textit{E. Coli}, Fruit Fly, and Human datasets sequenced using the latest nanopore chemistry (i.e., R10.4.1) show that \mech can be used to consistently improve the accuracy of raw signal analysis approaches that use a segmentation algorithm, such as raw signal mapping, without substantially increasing their computational resource requirements.

\mech makes the following key contributions:

\begin{itemize}
    \item We propose \mech, the first tool that corrects nanopore event sequences by learning from error-free nanopore event sequences.
    \item We show \mech's parameters can be optimized to utilize its error correction benefits with various segmentation algorithms and downstream analyses.
    \item We show that \mech reduces the parameter optimization requirement for newer chemistries, as segmentation parameters designed for an older nanopore chemistry (R9.4) can be used with a newer chemistry (R10.4.1) without requiring parameter optimization, by correcting errors with \mech.
    \item We show that using \mech-corrected events consistently improves the accuracy of the state-of-the-art raw signal analysis tool, RawHash2.
    \item We show that \mech adds minimal computational overhead, accounting for less than 1\% of the total read mapping runtime for larger genomes.
    \item We show that \mech can complement the commonly used homopolymer compression (HPC) mechanism, where applying a \mech configuration tuned for use with HPC consistently improves accuracy.
    \item We identify new directions for improving raw signal analyses, as well as challenges to overcome.
    \item We provide the open source code for training and testing \mech at \release.
\end{itemize}

\section{Methods} \label{xyz:sec:methods}
\subsection{Overview}

\begin{figure*}[tbh]
  \centering
  \includegraphics[
  width=0.7\linewidth,
  trim=2.2cm 2cm 0.7cm 0.5cm,
    clip
  ]{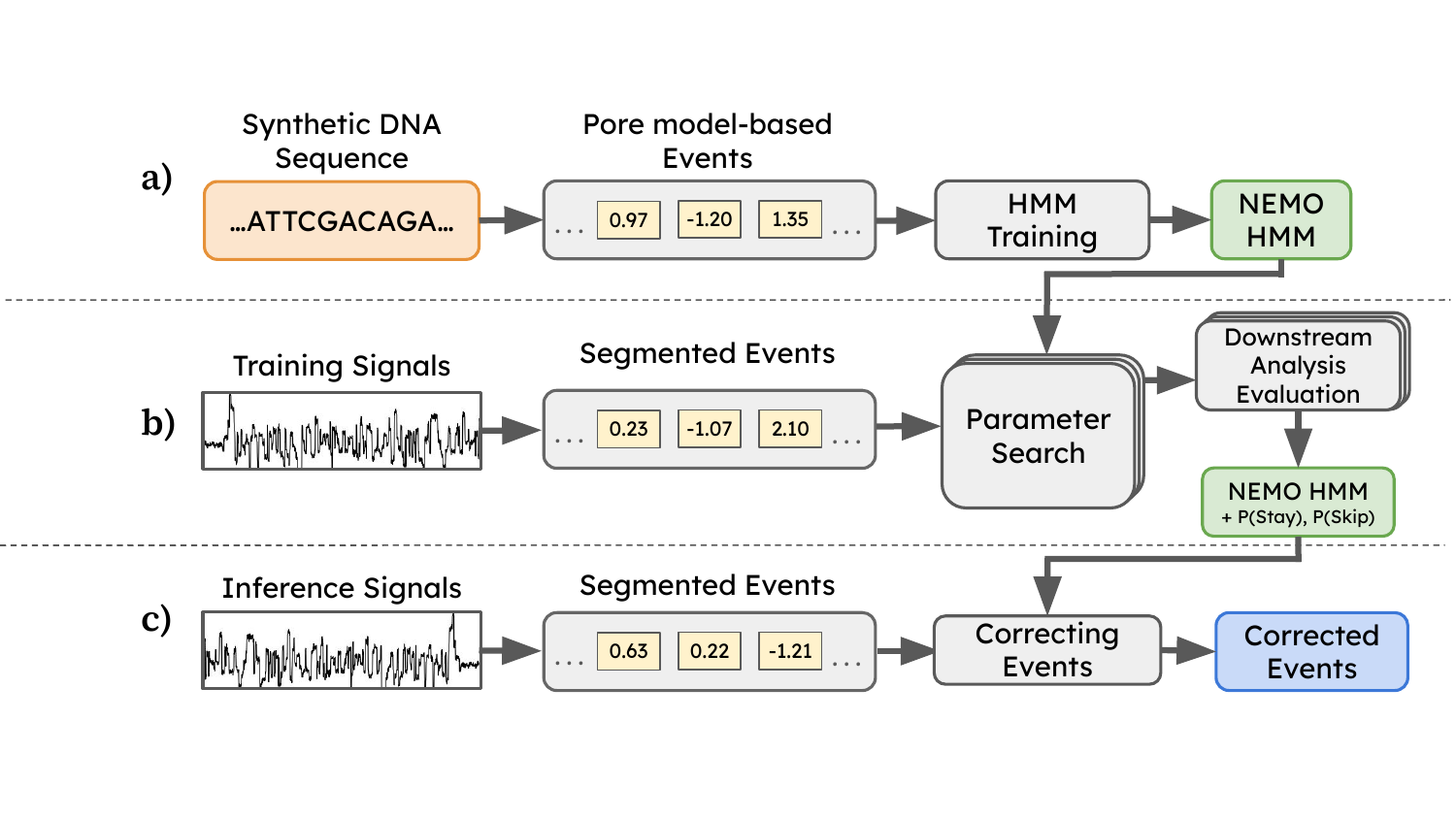}
  \caption{Overview of \mech.}
  \label{rs:methfig:cern_overview}
\end{figure*}

\mech is a mechanism to correct the errors in raw nanopore events (i.e., segmented raw nanopore signals). To achieve this, CRANE trains and uses a Hidden Markov Model (HMM). To train, it learns from both error-free and erroneous segmented signals. To correct (inference), it takes erroneous raw nanopore events as input to the trained HMM and outputs the corrected sequence of raw nanopore events, as shown in Figure~\ref{rs:methfig:cern_overview}.

\mech corrects raw nanopore events in three key steps.
First, \mech builds a base HMM that can effectively model nanopore events without being affected by segmentation errors or signal noise, or biased toward a specific DNA sequence (\circled{a}). To this end, the HMM is trained on synthetic, error-free event sequences using the Baum-Welch (BW) algorithm, which produces a sparsely connected HMM. We refer to the resulting models as nanopore event modeling HMMs (\textbf{NEMO-HMMs}).
Second, to optimize \mech's correction accuracy for use in a specific raw signal analysis pipeline (\circled{b}), \mech sets fixed transition probabilities within the trained HMM to profile certain error types. The optimal values of these fixed probabilities for a given pipeline are found via an empirical parameter search.
Third, to correct errors in a given sequence of input events during inference (\circled{c}), \mech runs a modified Viterbi algorithm using the trained NEMO-HMM and interprets the resulting state path to 1)~identify and remove oversegmentation errors and 2)~reduce noise in the event values.

\subsection{HMM Training}

\mech relies on HMMs because they can efficiently model the relationship between a nucleotide sequence passing through a nanopore and the corresponding electrical signals that the nanopore measures. An HMM is defined by a set of possible states and transition probabilities between each state. This is useful for modeling a nanopore because each HMM state can represent the nucleotide subsequence of length $k$, called a \textit{k-mer}, physically inside the nanopore at a given time. A k-mer in a nanopore is dependent on the k-mers which come immediately before and after it, just how an HMM state may transition to and from a limited number of states. An HMM generates an observed output, called an \textit{emission}, according to its current state. \mech uses HMMs with state-specific Gaussian distributions to model these emissions, with each distribution parameterized by a state-specific mean and variance. Hence, the model is referred to as a \textit{Gaussian HMM}. 

\mech can utilize HMMs with varying numbers of states. Early works which use HMMs for nanopore signal analysis generally use one state for each possible k-mer inside the nanopore~\cite{david_nanocall_2017}. For example, an HMM for the R10.4.1 nanopore chemistry would require $4^9=262144$ states. While this does enable more accurate analysis, it greatly increases the runtime of algorithms on the HMM. Since \mech HMMs have much less states than there are k-mers, there is no explicit mapping between states and k-mers. Instead, each state can be thought of as a clustering of k-mers learned during training.

\head{Initializing the HMM}
When training an HMM, good initial parameter estimates are important for proper convergence~\cite{rabiner-1989}. We improve training behavior by initializing HMM parameters using estimates derived from a list which contains the expected signal value for each possible k-mer inside the nanopore, called a \textit{pore model}~\cite{sam_kovaka_uncalled4_2025, firtina_rawhash_2023}.

The initial parameters for an HMM with $S$ states are obtained by modeling the distribution of all expected signal values from a pore model as a mixture of $S$ Gaussian distributions. This is done by fitting a \textit{Gaussian Mixture Model} (GMM) with $S$ components to the values in the pore model using expectation-maximization (EM)~\cite{dempster_1977}. To initialize the GMM component means, we cluster the pore model values into $S$ clusters via the k-means++ algorithm~\cite{kmeans_2007}, and take the cluster centers as the mean of each component. GMM component variances are uniformly initialized as the global variance of expected signal values divided by $S$. To initialize states of the HMM, \mech sets the mean and variance of each state's Gaussian distribution to the mean and variance of each GMM component. To initialize transitions, the HMM is given a uniform initial state distribution, and the transition probabilities are uniform with the exception that no state can transition to itself. This is because each time an event or observation is generated, it represents a new base of the nucleic acid molecule entering the nanopore, meaning the state should not be the same (with the exception of long runs of the same base).

\head{Training using Baum-Welch}
The HMM is trained on a synthetic nanopore event sequence, made by referencing a pore model. This procedure has three main advantages. First, it avoids the risk of training data being distorted by segmentation errors that are present in experimental data. Second, the learned distribution means of each state are accurate and noise-free, allowing for the HMM to detect some amount of noise in experimental data. Third, it prevents the model from becoming biased toward a specific genome or DNA sequence.

The event sequences are generated by creating a DNA sequence, then referencing the pore model for every k-mer in the sequence to find its expected event value. These values, taken in order, make up the training sequence. No noise is added to the event values. The DNA sequence used to generate the synthetic events is a \textit{de~Bruijn sequence} which contains every possible n-mer exactly once, where n is a tunable parameter.

\mech trains the HMM using the BW algorithm, which maximizes the probability of it generating the training sequence. For a sufficiently diverse sequence, this creates an HMM which can effectively model realistic, error-free nanopore event sequences. During training, only emission means, emission variances, and transition probabilities are updated. To improve training, we modify the BW algorithm in two ways. First, after a set amount of EM iterations, any transition with a sufficiently small probability is set to zero at the end of each subsequent iteration. This makes the HMM more sparsely connected, which speeds up the Viterbi algorithm used during error correction. The reason that many transitions tend to approach zero is that each k-mer can only transition from and to four other k-mers (using a DNA/RNA alphabet), putting a constraint on what transitions are possible. Second, the training sequence is optionally split up into $t$ sub-sequences to perform the expectation step in parallel. This significantly speeds up each EM iteration at the cost of a slight drop in accuracy, allowing large models to be trained in a reasonable amount of time.

\subsection{Parameter Search}

After training the NEMO-HMM on noise-free synthetic data, we adapt the model to real-world data for subsequent inference. We accomplish this by searching for the best correction parameters that reflect the observed error profile of the data (Figure~\ref{rs:methfig:cern_overview} \circled{b}). The optimal parameters may change depending on 1)~how the events are generated, and 2)~how the corrected events are used afterwards. As such, the parameter search needs to be performed for each combination of segmentation algorithm and downstream analysis the data goes through. We refer to each combination as an \textit{analysis pipeline}.

\mech searches over the number of states in the HMM, as well as the transition probabilities $P(\text{stay})$ and $P(\text{skip})$. $P(\text{stay})$ is the probability of a state transitioning to itself, which helps to model oversegmentation errors, also called \textit{stay errors}. $P(\text{skip})$ represents the chance of taking a zero-probability transition, which allows the HMM to take transitions which rarely or never occur in error-free data. These transitions may occur due to distortions such as large amounts of noise in the signal or undersegmentation errors. The transition parameters are applied to a base HMM in three steps. First, all self transition probabilities are initialized to $P(\text{stay})$. Second, all remaining zero-probability transitions are uniformly initialized such that all of them leaving a state sum to $P(\text{skip})$. Third, to ensure transition probability distributions add to 1, all original transitions in the HMM are multiplied by $1-P(\text{stay}) - P(\text{skip})$.

To find the optimal combination of parameters, \mech 1)~corrects the segmented training data by using the trained HMM for each P(stay), P(skip), and HMM size parameter configuration and 2)~uses the corrected sequence of events in a raw signal analysis tool (e.g., raw signal mapping with RawHash2~\cite{firtina_rawhash2_2024}) to identify the impact of the parameter choice in the accuracy of the analysis tool.

To iterate over parameter configurations, a grid search over combinations of $P(\text{stay})$ and $P(\text{skip})$ is performed for each number of states tested. After each grid point is evaluated using the raw signal analysis tool, the highest performing combination is then used as the starting point of a hill-climbing search. \mech tests increasing and decreasing both $P(\text{stay})$ and $P(\text{skip})$ by set step sizes. If \mech finds a new optimal set of parameters, it continues from the new optimal point. If a new optimum is not found, then the above process is repeated with a smaller step size. This is done until the hill-climb with the smallest step size converges.

\subsection{Correcting Events} \label{sec:correctionmethods}

\begin{figure}[tbh]
  \centering
  \includegraphics[
  width=\columnwidth,
  trim=1cm 0.5cm 8cm 2cm,
    clip
  ]{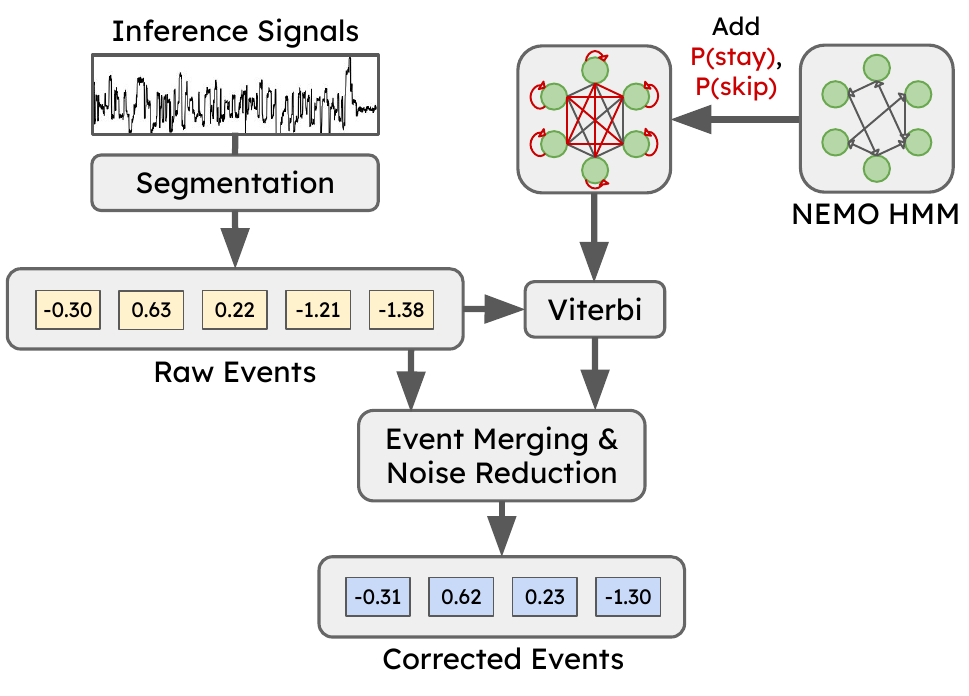}
  \caption{Error correction pipeline in \mech. The Viterbi path of states through the trained NEMO-HMM is used to identify and correct erroneous events.}
  \label{rs:methfig:correct_events}
\end{figure}

Once the best parameter configuration has been found, \mech corrects errors in event sequences by estimating the underlying statistical process for a given observation of events. \mech uses the Viterbi algorithm to decode the most likely sequence of states through the HMM to generate the sequence of events, which is then used to correct errors in two ways as shown in Figure~\ref{rs:methfig:correct_events}: 1)~removing oversegmentation errors and 2)~reducing noise in event values. It is possible to run \mech with both mechanisms, or to only use one of the two, allowing \mech to be effectively implemented in analysis pipelines where either oversegmentation or event noise is the dominant source of error.

\head{Efficient Inference via a Modified Viterbi Algorithm} \label{xyz:subsec:met_viterbi}
During training, the HMM becomes sparsely connected due to transition trimming, which allows the Viterbi algorithm to run efficiently by considering fewer transitions at each timestep. However, \mech reintroduces all previously removed transitions using the $P(\text{stay})$ and $P(\text{skip})$ parameters, causing the HMM to become densely connected again. To nevertheless exploit the efficient sparse structure learned during training while also retaining the non-sparse transitions found from the parameter search on experimental data, \mech uses a modified Viterbi algorithm: \mech tracks 1)~transitions belonging to the original sparse structure (\emph{sparse transitions}) and 2)~transitions pruned during training but reintroduced when loading the HMM according to the $P(\text{stay})$ and $P(\text{skip})$ parameters (\emph{non-sparse transitions}). At each timestep of the modified Viterbi algorithm, \mech considers all sparse transitions, but only the non-sparse transitions which are either 1)~self-transitions or 2)~transitions leaving the state with the largest log-likelihood at the previous timestep. This heuristic allows \mech to perform inference on the underlying sparse HMM while only considering two additional predecessors per state at each timestep. This substantially reduces the computational cost compared to the standard Viterbi algorithm while producing nearly identical results.

\head{Removing Stay Errors}
To identify stay errors in the sequence of event values, \mech considers the sequence of decoded HMM states produced by the Viterbi algorithm. A state which takes a transition to itself (i.e., self-transition) one or more times in a row is interpreted as the corresponding events being duplicated or oversegmented, since the path remains in the same state for multiple events. To correct such a stay error, \mech replaces these oversegmented events with a single event whose value is the average of those events.

\head{Reducing Noise in Event Values}
After removing oversegmentation errors, \mech reduces noise in the remaining event values. Here, we define noise as the difference between a sequenced k-mer's measured event value and its expected event value. To this end, \mech computes the difference between each event value and the emission mean of the HMM state it is aligned with in the Viterbi path. These differences represent how far each event deviates from the model's expectation. Due to the relatively small number of states (e.g. 128 states) in the HMM compared to the number of possible 9-mers (i.e., $4^9$ k-mers), these differences are not expected to be zero. Therefore, to detect local noise, \mech takes a windowed average of the differences between event values and state emission means. This windowed average is subsequently subtracted from the central event value in the middle of the window to produce the noise-adjusted sequence. This step is only performed \textit{after} stay errors have been removed. The windowed approach recognizes larger scale differences in sequential event values that may be represented as skewed signal values, and thus accounts for noise in the sequencing method. The size of the window is a tunable parameter.

\section{Results} \label{xyz:sec:results}
\subsection{Evaluation Methodology} \label{xyz:subsec:evaluation} 

We evaluate \mech by integrating it into the state-of-the-art raw signal mapping tool, RawHash2~\cite{firtina_rawhash2_2024}. RawHash2 performs segmentation, sketching, seeding, and chaining on raw nanopore events to map raw signals to a reference genome. We measure the impact of \mech on the accuracy (i.e., F1 score) and runtime of 1)~RawHash2 read mapping \textit{with} homopolymer compression (HPC), and 2)~RawHash2 read mapping \textit{without} HPC.

\head{Segmentation Algorithms}
We evaluate \mech with three segmentation configurations.
The first two are based on the t-test segmentation algorithm in Scrappie~\cite{oxford_nanopore_technologies_scrappie_2019}. Scrappie's segmentation algorithm performs a rolling Welch's t-test with windows of different lengths to identify points in the raw signal to detect significant changes. The first segmentation configuration uses the t-test parameters optimized for the R9.4 chemistry, which we call Scrappie (R9.4). The second segmentation configuration uses the parameters optimized for the R10.4.1 chemistry, called Scrappie (R10.4.1). The third segmentation algorithm is a deep learning-based tool, called Campolina. Campolina provides more accurate but slower segmentation than Scrappie~\cite{bakic_campolina_2026}. We integrate all of these segmentation algorithms within RawHash2 to perform end-to-end raw signal mapping.

\head{Training Configuration} HMMs are trained using the Baum-Welch (BW) algorithm on the expected R10.4.1 nanopore events of the sequence which contains every 11-mer exactly once, called the \textit{de Bruijn} sequence. After the first 100 EM iterations of BW, any transition with probability below 0.001 is set to zero at the end of each subsequent iteration. Each HMM is trained using 32 threads ($t=32$) by splitting the training sequence into 32 sub-sequences for parallelization. The pore model we reference for converting DNA sequences to expected event sequences is the \texttt{dna\_r10.4.1\_400bps\_9mer} model generated by Uncalled4~\cite{sam_kovaka_uncalled4_2025}. This pore model describes the expected signal value for every possible 9-mer.

\head{Parameter Search} We train eight HMMs, containing 8, 16, 32, 64, 96, 128, 192, and 256 states, respectively, to determine the optimal HMM size for each configuration of RawHash2 we test with. We do not test HMMs larger than 256 states due to increasing computational costs. Every parameter search is performed using each HMM size. For each of the three segmentation algorithms, we run two parameter searches: one where reads are mapped by RawHash2 with HPC off, and another where reads are mapped by RawHash2 with HPC on. We refer to \mech configurations optimized for use with HPC on as \mech-H. The set of reads corrected and mapped during the parameter searches is 12,000 reads from the \emph{E.~coli} sequencing data. This results in six sets of optimal $P(\text{stay})$ and $P(\text{skip})$ parameters for correcting Scrappie (R9.4), Scrappie (R10.4.1), and Campolina events, each with HPC on and off. To find the number of HMM states which performs the best for each RawHash2 configuration, the parameter searches are repeated for all eight trained HMMs of differing sizes. We find that 128 states is optimal when reads will be mapped \textit{without} HPC, and 16 states is optimal when reads will be mapped \textit{with} HPC.

To show the efficacy of the standalone stay error removal and noise reducing mechanisms of \mech, detailed in Section \ref{sec:correctionmethods}, we perform two supplementary sets of parameter searches across all eight HMM sizes, three datasets, and three segmentation algorithms. In the first, we find the optimal parameters when \mech only removes stay errors, and does not remove noise. In the second, \mech only removes noise, and does not remove stay errors. Both are evaluated using RawHash2 with HPC off. Stay error removal and noise reduction can be turned off by passing the \texttt{-{}-no-stay-removal} and \texttt{-{}-no-noise-removal} flags to \mech, respectively.

To identify the P(stay) and P(skip) parameters that maximize the accuracy of each analysis pipeline on the training data, we performed 1)~a grid search with all possible pairs of the values $\{0, 0.025, 0.05, 0.1,0.2,0.3\}$, and 2)~ a hill-climb with three step sizes: $\{0.02, 0.01, 0.005\}$.
The full results of all parameter sweeps can be found in Supplementary Tables \ref{tab:no_hpc_parameter_search} and \ref{tab:hpc_parameter_search}.

\head{Datasets} We utilize four datasets of raw nanopore reads sequenced using the R10.4.1 chemistry, as shown in Table~\ref{xyz:tab:dataset}: D0 (\emph{E.~coli}), D1 (\emph{E.~coli}), D2 (\emph{D.~melanogaster}, or Fruit Fly), and D3 (\emph{H.~sapiens}, or Human).
D0 is used for the parameter search, while D1, D2, and D3 are used solely for evaluation. Although D0 and D1 contain reads generated from the same experiment, they share no reads in common.

For each evaluation dataset, we evaluate the mapping of 10,000 reads before and after error correction with \mech, with and without HPC.
 
\begin{table*}[htb]
\centering
\caption{Details of datasets used in our evaluation.}
\begin{tabular}{@{}clllrrll@{}}
\toprule
& \textbf{Organism}
& \textbf{Chemistry}
& \textbf{Flow Cell}
& \textbf{Reads}
& \textbf{Avg. Length}
& \textbf{Basecaller}
& \textbf{Data Source} \\
\midrule
D0 & \emph{E.~coli} CFT073
   & R10.4.1 e8.2
   & FLO-MIN114
   & 12{,}000
   & 794
   & Dorado SUP v1.4.0
   & \cite{hall_benchmarking_2024} \\
\midrule
D1 & \emph{E.~coli} CFT073
   & R10.4.1 e8.2
   & FLO-MIN114
   & 10{,}000
   & 764
   & Dorado SUP v1.4.0
   & \cite{hall_benchmarking_2024} \\
\midrule
D2 & \emph{D.~melanogaster}
   & R10.4.1 e8.2
   & FLO-MIN114
   & 10{,}000
   & 5{,}998
   & Dorado SUP v0.9.2
   & \cite{ont_opendata_dmelanogaster} \\
\midrule
D3 & \emph{H.~sapiens} (HG002)
   & R10.4.1 e8.2
   & FLO-PRO114M
   & 10{,}000
   & 16{,}943
   & Dorado SUP v1.4.0
   & \cite{ont_opendata_giab2023} \\
\bottomrule
\multicolumn{8}{l}{\footnotesize All datasets use R10.4.1 e8.2 nanopore chemistry at 400~bases/sec translocation speed.} \\
\multicolumn{8}{l}{\footnotesize D0, D1 and D3 are sampled at 5{,}000~Hz; D2 at 4{,}000~Hz.} \\
\multicolumn{8}{l}{\footnotesize Each dataset contains a subsample of raw nanopore reads from the original sequencing run.} \\
\end{tabular}

\label{xyz:tab:dataset}
\end{table*}

\head{Ground Truth}
To generate the ground truth read mappings, we basecall each read using Dorado, then map the basecalled reads using minimap2. To generate the number of true positives (TP), false positives (FP), true negatives (TN), and false negatives (FN), we use the \texttt{pafstats.py} script used in RawHash2 to compare RawHash2 read mappings and the minimap2 read mappings~\cite{firtina_rawhash2_2024}. We modify \texttt{pafstats.py} such that it reports runtime for \textit{all reads processed}, as opposed to only reads which are mapped. We report precision ($P = TP / (TP + FP)$), recall ($R = TP / (TP + FN)$), and F1 score ($F1 = 2PR / (P + R)$) for each read mapping. Supplementary Table~\ref{tab:supp_versions} provides the details of the tools and their versions.

\head{Evaluation Setup} We evaluate four RawHash2 read mapping configurations: 1)~without HPC and without \mech, 2)~without HPC and with \mech, 3)~with HPC and without \mech, 4)~with HPC and with \mech (\mech-H). Table~\ref{tab:paramresults1} provides the configuration details of CRANE and CRANE-H for each evaluated segmentation algorithm: Scrappie (R9.4), Scrappie (R10.4.1), and Campolina. For all configurations, \mech is ran with the default noise removal window size of 20.
 
We run \mech error correction using 32 EPYC-7313 CPU cores with 64~GB of RAM. Each read mapping test uses RawHash2 with 32 EPYC-7313 cores and 512~GB of RAM. We use the \texttt{r10} preset of RawHash2 for the D1 and D2 datasets and the \texttt{r10fast} preset for the D3 dataset. Events are loaded using the \texttt{-{}-events-file} flag for both corrected and uncorrected events. We disable the built-in HPC using \texttt{-{}-sig-diff~-1} when evaluating configurations without HPC. When measuring runtimes, we measure only the \mech error correction and RawHash2 read mapping steps. The per-read runtimes for \mech and RawHash2 are the time it takes for a single thread to process one read for both tools.
 
We provide the parameter settings and versions for each tool in Supplementary Tables~\ref{tab:supp_parameters} and~\ref{tab:supp_versions} respectively. We provide the scripts to fully reproduce our results on the GitHub repository at \release.

\subsection{Parameter Search}

Table \ref{tab:paramresults1} shows the parameters found for \mech and \mech-H when correcting each segmentation algorithm. The parameter search for \mech was guided using the F1 scores of RawHash2 without HPC, and the search for \mech-H was guided using the F1 scores of RawHash2 with HPC. We report the \mech parameters for the 128-state HMM, and the \mech-H parameters for the 16-state HMM. Extended results reporting the parameter search results for HMMs of other sizes can be found in Supplementary Tables \ref{tab:no_hpc_parameter_search}, \ref{tab:hpc_parameter_search}. We make two observations.

First, the found $P(\text{stay})$ parameters are much lower when the corrected reads are mapped with HPC turned on. We believe this is because HPC removes a large portion of stay errors (i.e., oversegmentation errors), since events that are produced from the same oversegmentation error generally have similar values to each other. As such, \mech may have learned to be conservative when removing stay errors because HPC will remove the majority that remain.

Second, the number of states used in the HMM is much lower when optimizing for RawHash2 with HPC on. We chose 128 states for \mech, and 16 states for \mech-H because they yield the highest general performance when mapped with RawHash2 with HPC off, and HPC on, respectively. The final F1-scores of parameter searches across HMM sizes show that larger HMMs (96+ states) perform better when reads are mapped with HPC off, and smaller HMMs (16 states) perform better when reads are mapped with HPC on.

To investigate the discrepancy in HMM size, we perform two more sets of parameter searches, both of which are guided by RawHash2 with HPC off. The first parameter search is performed by \mech with only stay removal. The second search is performed with only noise removal. The full results for these searches are reported in Supplementary Table \ref{tab:ablation_parameter_search}. These searches show two important insights. First, noise reduction performs the best with 16 states, while stay error removal performs the best in the 96-256 state range. This explains why \mech-H performs best with 16 states; when HPC is on, \mech-H does not need to remove stay errors as aggressively, so it can optimize for noise removal. Second, both the stay error removal and noise reduction functions of \mech increase the accuracy of RawHash2 read mappings when applied independently.

\begin{table}[h]
\centering
\caption{Parameter search results across segmentation algorithms, with and without HPC}
\label{tab:paramresults1}
\begin{tabular}{@{}llcc@{}}
\toprule
\textbf{Segmentation} & \textbf{Parameter} & \textbf{\mech} & \textbf{\mech-H} \\
\midrule
\multirow{3}{*}{Scrappie (R9.4)}
& HMM States   & 128 & 16 \\
    & $P(\text{stay})$   & 0.3 & 0.2 \\
    & $P(\text{skip})$  & 0.02 & 0 \\
\midrule
\multirow{3}{*}{Scrappie (R10.4.1)}
& HMM States   & 128 & 16 \\
    & $P(\text{stay})$   & 0.05 & 0.01 \\
    & $P(\text{skip})$  & 0.02 & 0  \\
\midrule
\multirow{3}{*}{Campolina}
& HMM States   & 128 & 16 \\
    & $P(\text{stay})$   & 0.035 & 0 \\
    & $P(\text{skip})$  & 0.01 & 0.015  \\
\bottomrule
\end{tabular}

\end{table}

\subsection{Read Mapping Accuracy} \label{cern:subsec:results}
 
\head{\mech vs. Baseline without HPC}
Table~\ref{tab:results1} shows the effect of \mech on the F1 scores of RawHash2 read mapping without HPC across three segmentation algorithms on the D1, D2, and D3 datasets. Extended results reporting F1 score, precision, and recall across all tested configurations are available in Supplementary Tables~\ref{tab:full-results-D1},~\ref{tab:full-results-D2}, and~\ref{tab:full-results-D3}. We make three key observations.

\begin{table}[h]
\centering
\caption{F1 scores of RawHash2 read mapping without HPC across multiple datasets, segmentation algorithms, with and without \mech error correction.}
\label{tab:results1}
\begin{tabular}{@{}llcc@{}}
\toprule
\textbf{Dataset} & \textbf{Segmentation} & \textbf{Baseline} & \textbf{\mech} \\
\midrule
\multirow{3}{*}{D1 (\textit{E.~coli})}
    & Scrappie (R9.4)   & 0.177 & \textbf{0.726} \\
    & Scrappie (R10.4.1)  & 0.611 & \textbf{0.652} \\
    & Campolina    & 0.810 & \textbf{0.821} \\
\midrule
\multirow{3}{*}{D2 (Fruit Fly)}
    & Scrappie (R9.4)   & 0.321 & \textbf{0.840} \\
    & Scrappie (R10.4.1)  & 0.744 & \textbf{0.758} \\
    & Campolina    & 0.858 & \textbf{0.862} \\
\midrule
\multirow{3}{*}{D3 (Human)}
    & Scrappie (R9.4)   & 0.001 & \textbf{0.774} \\
    & Scrappie (R10.4.1)  & 0.682 & \textbf{0.755} \\
    & Campolina    & 0.853 & \textbf{0.856} \\
\bottomrule
\end{tabular}

\end{table}

First, \mech consistently improves the F1 score over the baseline across all experiments. \mech provides especially large improvements when paired with Scrappie (R9.4). We believe this is for two reasons: 1)~Scrappie (R9.4) tends to heavily oversegment R10.4.1 signals, and one of the primary ways \mech corrects events is by identifying and removing these oversegmentation errors. 2)~Scrappie (R9.4) is designed for an older nanopore chemistry, meaning any analysis tool which uses Scrappie (R9.4) on signals generated by an R10.4.1 nanopore without error removal (e.g. \mech or HPC) is expected to perform poorly. More detailed results on the effects of \mech and HPC can be found in Supplementary Tables~\ref{tab:hpc-delta} and \ref{tab:correction-delta}.

Second, \mech-corrected Scrappie (R9.4) events achieve substantially higher accuracy than uncorrected Scrappie (R10.4.1) events across all datasets (e.g. 0.774 vs.\ 0.682 on D3). On all datasets, \mech-corrected Scrappie (R9.4) events even outperform \mech-corrected Scrappie (R10.4.1) events. Since Scrappie (R9.4) heavily oversegments the signal, we believe this provides \mech with more information about the underlying signal structure than Scrappie (R10.4.1), allowing \mech to perform better as it excels at merging oversegmented events. \textbf{This result demonstrates that \mech can reduce the dependency of raw signal analysis on chemistry-specific segmentation parameter optimization, as segmentation parameters designed for an older chemistry (R9.4) can be effectively used with a newer chemistry (R10.4.1) when corrected by \mech.}

Third, \mech improves accuracy for both t-test-based (Scrappie (R9.4), Scrappie (R10.4.1)) and deep learning-based (Campolina) segmentation algorithms. The F1 scores with Campolina are substantially higher than those with the t-test-based segmenters across all configurations, since Campolina provides more accurate segmentation with fewer over- and undersegmentation errors. Even in this case, \mech still provides improvements, showing that \mech can benefit raw signal analysis even when the underlying segmentation algorithm is already highly accurate. Supplementary Table~\ref{tab:best-f1} also shows that the highest F1 score observed for each dataset is achieved by using Campolina in combination with \mech. 

\head{\mech-H vs. Baseline with HPC}
Table~\ref{tab:results2} shows the effect of \mech-H on the F1 scores of RawHash2 read mapping with HPC across all segmentation algorithms and datasets. We make three key observations.

\begin{table}[h]
\centering
\caption{F1 scores of RawHash2 read mapping with HPC across multiple datasets, segmentation algorithms, with and without \mech error correction.}
\label{tab:results2}
\begin{tabular}{@{}llcc@{}}
\toprule
\textbf{Dataset} & \textbf{Segmentation} & \textbf{\makecell{Baseline \\ + HPC}} & \textbf{\makecell{\mech-H \\ + HPC}} \\
\midrule
\multirow{3}{*}{D1 (\textit{E.~coli})}
    & Scrappie (R9.4)   & 0.697 & \textbf{0.823} \\
    & Scrappie (R10.4.1)  & 0.776 & \textbf{0.797} \\
    & Campolina    & 0.876 & \textbf{0.885} \\
\midrule
\multirow{3}{*}{D2 (Fruit Fly)}
    & Scrappie (R9.4)   & 0.798 & \textbf{0.883} \\
    & Scrappie (R10.4.1)  & 0.834 & \textbf{0.846} \\
    & Campolina    & 0.882 & \textbf{0.892} \\
\midrule
\multirow{3}{*}{D3 (Human)}
    & Scrappie (R9.4)   & 0.044 & \textbf{0.807} \\
    & Scrappie (R10.4.1)  & 0.787 & \textbf{0.803} \\
    & Campolina    & 0.875 & \textbf{0.882} \\
\bottomrule
\end{tabular}

\end{table}

First, applying \mech-H in combination with HPC consistently increases the F1 score compared to HPC alone across all combinations. This demonstrates the ability of \mech and HPC to complement each other for a consistent improvement in accuracy.

Second, including HPC increases F1 scores across all tests, and in some cases provides a larger increase in performance than \mech (e.g., from 0.611 to 0.776 for D1 with ScrappieR10). We believe this is because 1)~parameters and heuristics in RawHash2 are optimized for use with HPC, and 2)~HPC acts on both the raw reads and the reference, while \mech only acts on the reads. This distinction shows that \mech and HPC are fundamentally different approaches: \mech performs error correction on the reads alone with no knowledge of a reference, whereas HPC transforms both the reads and the reference into representations that are inherently more similar.

\subsection{Computational Overhead} \label{cern:subsec:runtime}
 
\begin{figure}[tbh]
  \centering
  \includegraphics[
  width=0.95\columnwidth,
    clip,
    trim=0mm 0mm 0mm 0mm
  ]{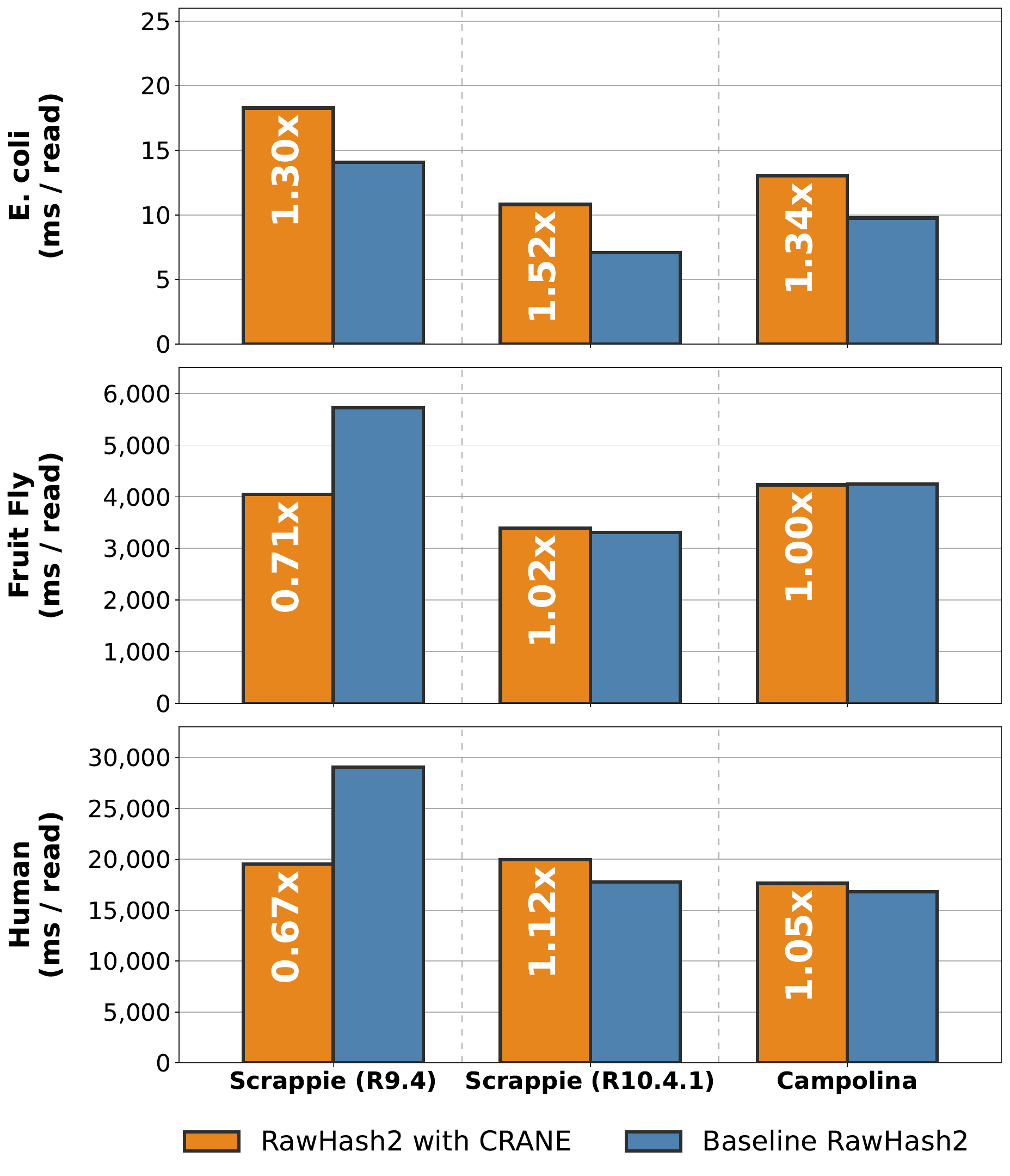}
  \caption{Runtime of RawHash2 with and without \mech error correction when HPC is disabled. Baseline denotes the original RawHash2 pipeline using the corresponding segmenter without \mech error correction. The labels show how many times longer the combined read mapping and error correction pipeline takes compared to the corresponding baseline.}
  \label{rs:resultsfig:runtime_events}
\end{figure}

\head{Runtime of \mech and RawHash2 without HPC} Figure~\ref{rs:resultsfig:runtime_events} shows the combined runtime of \mech and RawHash2 read mapping without HPC, as well as the runtime of RawHash2 without \mech and without HPC across all datasets and segmentation algorithms. The exact runtimes of RawHash2 and \mech are reported in Table~\ref{tab:runtimes_hpc_off}, and Supplementary Tables \ref{tab:full-results-D1}, \ref{tab:full-results-D2}, and \ref{tab:full-results-D3}. We make two key observations.

\begin{table}[h]
\centering
\caption{Per-read runtimes (ms) of 1)~only \mech and 2)~RawHash2 without HPC, with (\mech + RH2) and without (Baseline RH2) \mech error correction. The percentages in \mech + RH2 show the percentage of runtime attributed to \mech.}
\resizebox{\linewidth}{!}{
\begin{tabular}{clrrr}
\toprule
\textbf{Dataset} & \textbf{Segmentation} & \textbf{\mech} &\textbf{ \makecell{\mech\\+ RH2}} & \textbf{\makecell{Baseline\\RH2}} \\
\midrule
\multirow{3}{*}{\shortstack{D1\\ \textit{(E.\ coli)}}} & Campolina & 2.99  & 13.03 (22.95\%) & 9.75 \\
 & Scrappie (R9.4) & 8.00  & 18.27 (43.79\%) & 14.07  \\
 & Scrappie (R10.4.1) & 3.43  & 10.81 (31.73\%) & 7.09 \\
\midrule
\multirow{3}{*}{\shortstack{D2\\ (Fruit Fly)}} & Campolina & 31.80 & 4,231.24 (0.75\%) & 4,248.04 \\
 & Scrappie (R9.4) & 49.43 & 4,046.91 (1.22\%) & 5,719.84  \\
 & Scrappie (R10.4.1) & 43.29 & 3,390.91 (1.28\%) & 3,312.10  \\
\midrule
\multirow{3}{*}{\shortstack{D3\\ (Human)}} & Campolina & 70.07 & 17,644.26 (0.40\%) & 16,791.35 \\
 & Scrappie (R9.4) & 150.74  & 19,516.55 (0.77\%) & 29,062.97 \\
 & Scrappie (R10.4.1) & 105.27 & 19,969.32 (0.53\%) & 17,770.87 \\
\bottomrule
\end{tabular}
}

\label{tab:runtimes_hpc_off}
\end{table}

First, \mech adds a very small computational overhead compared to the read mapping step, especially for larger genomes. For the D2 and D3 datasets, \mech correction accounts for less than 1\% of the total read mapping runtime. For the D1 dataset, the overhead is higher (22--44\%), which we attribute to the already very fast read mapping times for this small genome (below 30~ms per read on average).

Second, the runtime ratios between RawHash2 with \mech and baseline RawHash2 are substantially lower for Scrappie (R9.4)-segmented reads compared to other segmenters. Since Scrappie (R9.4) heavily oversegments the signal, \mech merges many events in each read, reducing the event sequence length which speeds up the mapping process. This brings the mapping runtime of Scrappie (R9.4)-segmented reads closer to that of Scrappie (R10.4.1) and Campolina segmented reads.

\begin{figure}[tbh]
  \centering
  \includegraphics[
  width=0.95\columnwidth,
    clip,
    trim=0mm 0mm 0mm 0mm
  ]{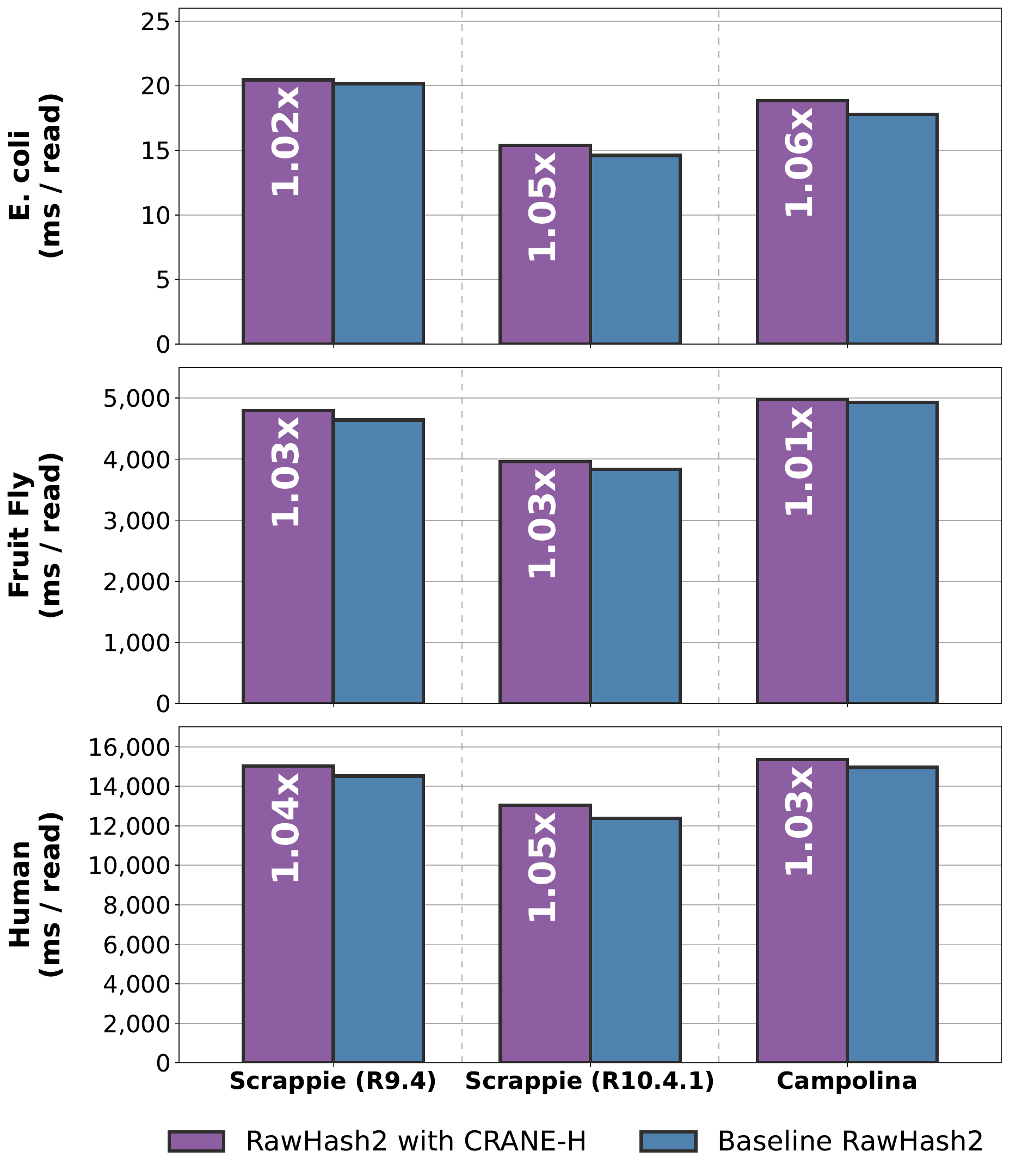}
  \caption{Runtime of RawHash2 with and without \mech-H error correction when HPC is enabled. Baseline denotes the original RawHash2 pipeline using the corresponding segmenter without \mech error correction. The labels show how many times longer the combined read mapping and error correction pipeline takes compared to the corresponding baseline.}
  \label{rs:resultsfig:runtime_events2}
\end{figure}

\head{Runtime of \mech-H and RawHash2 with HPC} Figure~\ref{rs:resultsfig:runtime_events2} shows the runtime of RawHash2 read mapping when using HPC in combination with \mech-H. The exact runtimes of RawHash2 and \mech-H are reported in Table \ref{tab:runtimes_hpc_on}, and Supplementary Tables \ref{tab:full-results-D1}, \ref{tab:full-results-D2}, and \ref{tab:full-results-D3}. We make two key observations.

\begin{table}[h]
\centering
\caption{Per-read runtimes (ms) of 1)~only \mech-H and 2)~RawHash2 with HPC, with (\mech-H + RH2) and without (Baseline RH2) \mech-H error correction. The percentages in \mech-H + RH2 show the percentage of runtime attributed to \mech-H.}
\resizebox{\linewidth}{!}{
\begin{tabular}{clrrr}
\toprule
\textbf{Dataset} & \textbf{Segmentation} & \textbf{\mech-H} & \textbf{\makecell{\mech-H\\+ RH2}} & \textbf{\makecell{Baseline\\RH2}} \\
\midrule
\multirow{3}{*}{\shortstack{D1\\ \textit{(E.\ coli)}}} & Campolina & 0.27 & 18.84 (1.43\%) & 17.79 \\
 & Scrappie (R9.4) & 0.43 & 20.46 (2.10\%) & 20.15 \\
 & Scrappie (R10.4.1) & 0.27 & 15.39 (1.75\%) & 14.60 \\
\midrule
\multirow{3}{*}{\shortstack{D2\\ (Fruit Fly)}} & Campolina & 2.26 & 4,975.73 (0.05\%) & 4,931.82 \\
 & Scrappie (R9.4) & 6.49 & 4,799.88 (0.14\%) & 4,640.95 \\
 & Scrappie (R10.4.1) & 3.14 & 3,956.69 (0.08\%) & 3,831.61 \\
\midrule
\multirow{3}{*}{\shortstack{D3\\ (Human)}} & Campolina & 6.97 & 15,358.02 (0.05\%) & 14,954.14 \\
 & Scrappie (R9.4) & 17.04 & 15,025.79 (0.11\%) & 14,511.13 \\
 & Scrappie (R10.4.1) & 11.72 & 13,034.23 (0.09\%) & 12,379.21 \\
\bottomrule
\end{tabular}

\label{tab:crane_overhead}
}

\label{tab:runtimes_hpc_on}
\end{table}

First, the computational overhead of \mech-H is near-negligible, particularly for the D2 and D3 datasets (e.g. 0.05\%-0.11\%). This is because \mech-H uses a 16-state HMM, greatly speeding up the Viterbi algorithm used during correction. This shows that \mech can improve the accuracy of a raw signal analysis pipeline while adding minimal computational overhead when used in combination with HPC.

Second, when used with HPC, \mech-H accounts for a smaller share of total runtime than the \mech configuration without HPC, because \mech-H uses a 16-state instead of a 128-state HMM. These results show that \mech has a flexible runtime which depends on the chosen HMM size. Nevertheless, both the 128-state and 16-state HMMs used in \mech and \mech-H, respectively, are practical for real-time raw signal analysis pipelines, as their computational cost is negligible relative to the downstream read mapping time for datasets of typical size.

We conclude that \mech provides consistent improvements in read mapping accuracy across different segmentation algorithms and datasets while adding minimal computational overhead to the raw signal analysis pipeline. 

\section{Discussion} \label{xyz:sec:discussion}
We discuss the benefits of \mech for raw signal analysis, its current limitations, and promising directions for future work.

\head{Benefits for Raw Signal Analysis}
The primary benefit of \mech is its ability to improve the accuracy of raw signal analysis tools that rely on segmentation algorithms. The motivation behind raw signal analysis is to perform analyses faster and with fewer computational resources than basecalling, enabling real-time decision-making and large-scale analyses. This efficiency comes at the cost of accuracy, as signal processing algorithms are inherently more error-prone than basecalling, which has become highly accurate with the release of the R10.4.1 nanopore chemistry~\cite{sanderson_accuracy_2024}. By correcting the errors that segmentation algorithms introduce, \mech can alleviate the impact of noisy and erroneous events on downstream raw signal analyses without requiring the computational overhead of basecalling.

Additionally, \mech can reduce the burden of chemistry-specific segmentation parameter optimization. Our results show that applying \mech to Scrappie (R9.4), a segmenter that severely oversegments R10.4.1 signals, results in read mapping accuracy that is consistently higher than applying \mech to Scrappie (R10.4.1) across all experiments.  This suggests that an effective strategy could be to apply \mech to a computationally inexpensive segmentation algorithm that tends to oversegment, rather than investing effort in optimizing segmentation parameters for each new chemistry. Such an approach can improve accuracy while saving the effort required for parameter tuning.

Lastly, \mech can be applied in a variety of raw signal analysis pipelines. The parameter search can be altered depending both on what segmentation algorithm is used, as well as what downstream task the \mech-corrected events are used for. For example, in this paper, we demonstrated that \mech can be optimized for three segmentation algorithms across two different configurations of RawHash2: HPC enabled, and HPC disabled. Other downstream tasks besides read mapping could be incorporated into the parameter search, making \mech a flexible framework that could be applied across raw signal analysis tools.

\head{Limitations}
There are two main limitations that \mech currently faces.
First, \mech does not correct undersegmentation errors. During our evaluations, we find that \mech struggles to model missing events as well as other complex segmentation errors that distort the signal more heavily than event duplication or noise which spans many events. In our initial efforts, we find that attempting to model these complex errors also significantly increases the error correction runtime. We expect that a mechanism capable of detecting and correcting such complex segmentation errors would likely require computational resources comparable to basecalling, which would reduce the practical benefits of raw signal analysis. However, it is possible that alternative model designs could better capture and correct these patterns.

Second, \mech does not explicitly account for nucleic acid modifications such as methylation. The synthetic DNA sequences used to train \mech contain only unmodified bases because 1)~the pore model files \mech relies on to generate expected event sequences mainly describe unmodified bases, and 2)~incorporating base modifications substantially expands the k-mer space. For example, considering a single modification type in R10.4.1 chemistry increases the number of possible 9-mers from $4^9 = 262{,}144$ to $5^9 = 1{,}953{,}125$. This expansion greatly increases the complexity of modeling nanopore events and would likely require a larger HMM with more states, increasing the runtime of \mech. It is possible that \mech does not need to explicitly model nucleic acid modifications. Because the HMM state-space is much smaller than the set of all possible k-mers, the model may learn the general patterns of nanopore signals rather than explicit k-mer identities. If signals generated by modified k-mers follow the same underlying patterns, then \mech may be able to accurately correct both modified and unmodified sequences.

\head{Future Work}
We identify promising directions for future work.
The area of correcting raw nanopore signals remains under-explored. \mech is the first mechanism that systematically learns from data to correct errors in event sequences. We believe there is an opportunity for tools that can detect sequences or regions of a sequence that are highly erroneous. Such detection could be used to flag high-error reads to prevent further processing, or to leverage Read Until technology~\cite{loose_real-time_2016} to eject a read during sequencing and begin generating a new one that is more likely to contain high-quality information.

While \mech addresses a key challenge in raw signal analysis by correcting segmentation errors, several opportunities remain to extend this work. Investigating new methods of error correction, addressing complex segmentation errors, and investigating the correction of modified bases has the potential to further improve the accuracy and scope of raw signal analyses.

\section{Conclusion} \label{xyz:sec:conclusion}
We introduce \mech, the \emph{first} mechanism that corrects raw nanopore event sequences by learning from error-free nanopore event sequences using Hidden Markov Models.
We find that \mech 1)~consistently improves the read mapping accuracy of the state-of-the-art raw signal analysis tool, RawHash2, across all tested segmentation algorithms and datasets, 2)~reduces the dependency of raw signal analysis on chemistry-specific segmentation parameter optimization, as segmentation parameters designed for an older nanopore chemistry (R9.4) can be effectively used with a newer chemistry (R10.4.1) when corrected by \mech, and 3)~adds minimal computational overhead.
We show that \mech-corrected events from a segmenter optimized for R9.4 chemistry (Scrappie (R9.4)) achieve substantially higher mapping accuracy on R10.4.1 data than uncorrected events from a segmenter optimized for R10.4.1 (Scrappie (R10.4.1)) across all datasets, demonstrating that \mech can alleviate the burden of re-optimizing segmentation parameters for each new nanopore chemistry. We also show that \mech improves accuracy for both lightweight statistical segmenters and more accurate deep learning-based segmenters, indicating that \mech provides general-purpose benefits regardless of the underlying segmentation approach.
We hope and believe that \mech enables future work. Correcting raw nanopore signals is an under-explored area, and \mech demonstrates that learned models of nanopore signals can systematically improve event quality. We believe this can inspire a new class of error correction mechanisms designed specifically for raw nanopore signals.

\section*{Acknowledgments}
We thank the STORM Research Group members for their feedback. STORM Research Group acknowledges the gifts from Advanced Micro Devices (AMD) and AMD University Program's AI \& HPC Cluster.

\bibliographystyle{IEEEtran}
{\bibliography{main}}
\onecolumn
\newcommand{\supptitle}[1]{%
  \begingroup
    \fontsize{17pt}{19pt}\selectfont
    \bfseries
    #1\par
  \endgroup
}
\setcounter{secnumdepth}{3}
\clearpage
\begin{center}
\supptitle{Supplementary Material for\\\ltitle}
\end{center}
\setcounter{section}{0}
\setcounter{equation}{0}
\setcounter{figure}{0}
\setcounter{table}{0}
\setcounter{page}{1}
\makeatletter
\renewcommand{\theequation}{S\arabic{equation}}
\renewcommand{\thetable}{S\arabic{table}}
\renewcommand{\thefigure}{S\arabic{figure}}
\renewcommand{\thesection}{\Alph{section}}
\renewcommand{\thesubsection}{\thesection.\arabic{subsection}}
\renewcommand{\thesubsubsection}{\thesubsection.\arabic{subsubsection}}

\newcommand{\TextUnderscore}{\rule{.4em}{.4pt}}
\section{Configuration}\label{rs:suppsec:configuration}
\subsection{Parameters} 

In Supplementary Table~\ref{tab:supp_parameters}, we list the parameters used for each tool and dataset. For minimap2~\cite{li_minimap2_2018}, we use the same parameter setting for all datasets. For the Dorado super-accurate (SUP) basecaller, we use the model trained for the corresponding data sampling frequency (i.e. 4 kHz or 5 kHz).

\begin{table}[tbh]
\centering
\caption{Parameters we use in our evaluation for each tool and dataset in mapping.}
\label{tab:supp_parameters}
\begin{tabular}{@{}lccc@{}}\toprule
\textbf{Tool} & \textbf{\emph{E. coli}} & \textbf{\emph{D. melanogaster}} & \textbf{\emph{H. sapiens}} \\\midrule
Minimap2          & -x map-ont & -x map-ont & -x map-ont\\\midrule
Dorado GPU (SUP) & \multicolumn{3}{c}{basecaller dna\_r10.4.1\_e8.2\_400bps\_sup@\textbf{v4.1.0}/\textbf{4.1.0}/\textbf{5.0.0}}\\\midrule
RawHash2 & -x r10 -w 0 & -x r10 -w 0 & -x r10fast -w 0 \\\midrule
\end{tabular}
\end{table}
\label{suppsubsec:parameters}
\subsection{Versions}\label{suppsubsec:versions}

Supplementary Table~\ref{tab:supp_versions} lists the version and the link to the corresponding versions of each tool we use in our experiments. Scripts to reproduce all experiments can be found on \release. We use Dorado 1.4.0 for D1 and D3 and 0.9.2 for D2 due to the differences in the library kit versions used when sequencing these datasets.

\begin{table}[tbh]
\centering
\caption{Versions of each tool and library.}
\label{tab:supp_versions}
\begin{tabular}{@{}lll@{}}\toprule
\textbf{Tool} & \textbf{Version} & \textbf{Link to the Source Code} \\\midrule
RawHash2 & 2.5 & \url{https://github.com/STORMgroup/RawHash2/commit/92366af}\\\midrule
Minimap2 & 2.24-r1122 & \url{https://github.com/lh3/minimap2/releases/tag/v2.24}\\\midrule
Dorado & 0.9.2 & \url{https://github.com/nanoporetech/dorado/releases/tag/v0.9.2}\\\midrule
Dorado & 1.4.0 & \url{https://github.com/nanoporetech/dorado/releases/tag/v1.4.0}\\\midrule
Uncalled4 & 4.1.0 & \url{https://github.com/skovaka/uncalled4/releases/tag/4.1.0}\\\midrule
\end{tabular}

\end{table}
\vspace{1em}

\section{Extended Results}\label{rs:suppsec:extended_results}

\vspace{1em}
\subsection{Best Configurations}\label{rs:suppsec:best_configuration}

In Supplementary Table~\ref{tab:best-f1} we report the best configuration for each dataset in terms of F1 scores.

\vspace{1em}

\begin{table}[htbp]
\centering
\caption{Best F1 configuration per dataset.}
\label{tab:best-f1}
\small
\begin{tabular}{l lcc rrrr}
\toprule
Dataset & Segmenter & HPC & Corrected & F1 & Precision & Recall & Time to Map (ms) \\
\midrule
  \textit{E.~coli} & Campolina & \ding{51} & \ding{51} & \textbf{0.885} & 0.993 & 0.799 & 18.57 \\
\midrule
  \textit{D.~melanogaster} & Campolina & \ding{51} & \ding{51} & \textbf{0.892} & 0.982 & 0.818 & 4,973.47 \\
\midrule
  \textit{H.~sapiens} & Campolina & \ding{51} & \ding{51} & \textbf{0.882} & 0.982 & 0.800 & 15,351.05 \\
\bottomrule
\end{tabular}
\end{table}

\clearpage
\subsection{Results by Dataset}\label{rs:suppsec:extended_results_datasets}

We show extended results for the datasets \textit{E. coli} (D1) in Supplementary Table~\ref{tab:full-results-D1}, \textit{D. melanogaster} (D2) in Supplementary Table~\ref{tab:full-results-D2}, and \textit{H. sapiens} (D3) in Supplementary Table~\ref{tab:full-results-D3}. For each configuration of segmentation method, HPC, and error correction we report mapping quality metrics in terms of F1 score, precision, recall and percentage of reads mapped, as well as performance in terms of average time spent per read.

\vspace{1em}

\begin{table}[htbp]
\centering
\caption{Full results for \textit{E.~coli} (D1).}
\label{tab:full-results-D1}
\small
\begin{tabular}{l cc rrrr}
\toprule
Segmenter & HPC & Corrected & F1 & Precision & Recall & Time to Map (ms) \\
\midrule
  Campolina & \ding{55} & \ding{51} & 0.821 & 0.991 & 0.701 & 10.04 \\
  Campolina & \ding{55} & \ding{55} & 0.810 & 0.987 & 0.687 & 9.75 \\
  Campolina & \ding{51} & \ding{51} & 0.885 & 0.993 & 0.799 & 18.57 \\
  Campolina & \ding{51} & \ding{55} & 0.876 & 0.992 & 0.785 & 17.79 \\
  Scrappie (R9.4) & \ding{55} & \ding{51} & 0.726 & 0.979 & 0.577 & 10.27 \\
  Scrappie (R9.4) & \ding{55} & \ding{55} & 0.177 & 0.743 & 0.100 & 14.07 \\
  Scrappie (R9.4) & \ding{51} & \ding{51} & 0.823 & 0.990 & 0.705 & 20.03 \\
  Scrappie (R9.4) & \ding{51} & \ding{55} & 0.697 & 0.974 & 0.543 & 20.15 \\
  Scrappie (R10.4.1) & \ding{55} & \ding{51} & 0.652 & 0.970 & 0.491 & 7.38 \\
  Scrappie (R10.4.1) & \ding{55} & \ding{55} & 0.611 & 0.952 & 0.450 & 7.09 \\
  Scrappie (R10.4.1) & \ding{51} & \ding{51} & 0.797 & 0.986 & 0.668 & 15.12 \\
  Scrappie (R10.4.1) & \ding{51} & \ding{55} & 0.776 & 0.985 & 0.640 & 14.60 \\
\bottomrule
\end{tabular}
\end{table}

\begin{table}[htbp]
\centering
\caption{Full results for \textit{D.~melanogaster} (D2).}
\label{tab:full-results-D2}
\small
\begin{tabular}{l cc rrrr}
\toprule
Segmenter & HPC & Corrected & F1 & Precision & Recall & Time to Map (ms) \\
\midrule
  Campolina & \ding{55} & \ding{51} & 0.862 & 0.974 & 0.773 & 4,199.44 \\
  Campolina & \ding{55} & \ding{55} & 0.858 & 0.958 & 0.777 & 4,248.04 \\
  Campolina & \ding{51} & \ding{51} & 0.892 & 0.982 & 0.818 & 4,973.47 \\
  Campolina & \ding{51} & \ding{55} & 0.882 & 0.982 & 0.801 & 4,931.82 \\
  Scrappie (R9.4) & \ding{55} & \ding{51} & 0.840 & 0.980 & 0.736 & 3,997.48 \\
  Scrappie (R9.4) & \ding{55} & \ding{55} & 0.321 & 0.593 & 0.220 & 5,719.84 \\
  Scrappie (R9.4) & \ding{51} & \ding{51} & 0.883 & 0.982 & 0.802 & 4,793.39 \\
  Scrappie (R9.4) & \ding{51} & \ding{55} & 0.798 & 0.979 & 0.673 & 4,640.95 \\
  Scrappie (R10.4.1) & \ding{55} & \ding{51} & 0.758 & 0.977 & 0.620 & 3,347.62 \\
  Scrappie (R10.4.1) & \ding{55} & \ding{55} & 0.744 & 0.977 & 0.601 & 3,312.10 \\
  Scrappie (R10.4.1) & \ding{51} & \ding{51} & 0.846 & 0.982 & 0.743 & 3,953.55 \\
  Scrappie (R10.4.1) & \ding{51} & \ding{55} & 0.834 & 0.984 & 0.724 & 3,831.61 \\
\bottomrule
\end{tabular}
\end{table}

\begin{table}[htbp]
\centering
\caption{Full results for \textit{H.~sapiens} (D3).}
\label{tab:full-results-D3}
\small
\begin{tabular}{l cc rrrr}
\toprule
Segmenter & HPC & Corrected & F1 & Precision & Recall & Time to Map (ms) \\
\midrule
  Campolina & \ding{55} & \ding{51} & 0.856 & 0.986 & 0.756 & 17,574.19 \\
  Campolina & \ding{55} & \ding{55} & 0.853 & 0.985 & 0.752 & 16,791.35 \\
  Campolina & \ding{51} & \ding{51} & 0.882 & 0.982 & 0.800 & 15,351.05 \\
  Campolina & \ding{51} & \ding{55} & 0.875 & 0.982 & 0.788 & 14,954.14 \\
  Scrappie (R9.4) & \ding{55} & \ding{51} & 0.774 & 0.988 & 0.636 & 19,365.81 \\
  Scrappie (R9.4) & \ding{55} & \ding{55} & 0.001 & 0.006 & 0.001 & 29,062.97 \\
  Scrappie (R9.4) & \ding{51} & \ding{51} & 0.807 & 0.988 & 0.682 & 15,008.75 \\
  Scrappie (R9.4) & \ding{51} & \ding{55} & 0.044 & 0.841 & 0.023 & 14,511.13 \\
  Scrappie (R10.4.1) & \ding{55} & \ding{51} & 0.755 & 0.980 & 0.614 & 19,864.05 \\
  Scrappie (R10.4.1) & \ding{55} & \ding{55} & 0.682 & 0.963 & 0.528 & 17,770.87 \\
  Scrappie (R10.4.1) & \ding{51} & \ding{51} & 0.803 & 0.985 & 0.678 & 13,022.51 \\
  Scrappie (R10.4.1) & \ding{51} & \ding{55} & 0.787 & 0.984 & 0.656 & 12,379.21 \\
\bottomrule
\end{tabular}
\end{table}

\clearpage
\subsection{Ablation Results for HPC}

In Supplementary Table~\ref{tab:hpc-delta} we report the effect of HPC on mapping quality and performance for each dataset and configuration. That is,
for each configuration of segmentation method and error correction
we report the difference between using and not using HPC. Improvements in a metric are displayed in teal, regressions in red.

\vspace{1em}

\begin{table}[htbp]
\centering
\caption{Effect of homopolymer compression (HPC): difference (HPC on $-$ HPC off).}
\label{tab:hpc-delta}
\small
\begin{tabular}{ll c rrrr}
\toprule
Dataset & Segmenter & Corrected & $\Delta$F1 & $\Delta$Precision & $\Delta$Recall & $\Delta$Time to Map (ms) \\
\midrule
  \textit{E.~coli} & Campolina & \ding{51} & \textcolor{teal}{+0.065} & \textcolor{teal}{+0.001} & \textcolor{teal}{+0.099} & \textcolor{red}{+8.53} \\
   & Campolina & \ding{55} & \textcolor{teal}{+0.066} & \textcolor{teal}{+0.005} & \textcolor{teal}{+0.098} & \textcolor{red}{+8.04} \\
   & Scrappie (R9.4) & \ding{51} & \textcolor{teal}{+0.097} & \textcolor{teal}{+0.010} & \textcolor{teal}{+0.128} & \textcolor{red}{+9.76} \\
   & Scrappie (R9.4) & \ding{55} & \textcolor{teal}{+0.521} & \textcolor{teal}{+0.231} & \textcolor{teal}{+0.443} & \textcolor{red}{+6.08} \\
   & Scrappie (R10.4.1) & \ding{51} & \textcolor{teal}{+0.144} & \textcolor{teal}{+0.016} & \textcolor{teal}{+0.177} & \textcolor{red}{+7.74} \\
   & Scrappie (R10.4.1) & \ding{55} & \textcolor{teal}{+0.165} & \textcolor{teal}{+0.032} & \textcolor{teal}{+0.191} & \textcolor{red}{+7.51} \\
\midrule
  \textit{D.~melanogaster} & Campolina & \ding{51} & \textcolor{teal}{+0.030} & \textcolor{teal}{+0.008} & \textcolor{teal}{+0.045} & \textcolor{red}{+774.03} \\
   & Campolina & \ding{55} & \textcolor{teal}{+0.024} & \textcolor{teal}{+0.024} & \textcolor{teal}{+0.024} & \textcolor{red}{+683.78} \\
   & Scrappie (R9.4) & \ding{51} & \textcolor{teal}{+0.043} & \textcolor{teal}{+0.002} & \textcolor{teal}{+0.067} & \textcolor{red}{+795.91} \\
   & Scrappie (R9.4) & \ding{55} & \textcolor{teal}{+0.477} & \textcolor{teal}{+0.386} & \textcolor{teal}{+0.453} & \textcolor{teal}{-1,078.89} \\
   & Scrappie (R10.4.1) & \ding{51} & \textcolor{teal}{+0.087} & \textcolor{teal}{+0.004} & \textcolor{teal}{+0.123} & \textcolor{red}{+605.93} \\
   & Scrappie (R10.4.1) & \ding{55} & \textcolor{teal}{+0.090} & \textcolor{teal}{+0.007} & \textcolor{teal}{+0.123} & \textcolor{red}{+519.51} \\
\midrule
  \textit{H.~sapiens} & Campolina & \ding{51} & \textcolor{teal}{+0.026} & \textcolor{red}{-0.004} & \textcolor{teal}{+0.044} & \textcolor{teal}{-2,223.14} \\
   & Campolina & \ding{55} & \textcolor{teal}{+0.022} & \textcolor{red}{-0.002} & \textcolor{teal}{+0.036} & \textcolor{teal}{-1,837.21} \\
   & Scrappie (R9.4) & \ding{51} & \textcolor{teal}{+0.033} & \textcolor{red}{-0.000} & \textcolor{teal}{+0.046} & \textcolor{teal}{-4,357.06} \\
   & Scrappie (R9.4) & \ding{55} & \textcolor{teal}{+0.043} & \textcolor{teal}{+0.834} & \textcolor{teal}{+0.022} & \textcolor{teal}{-14,551.84} \\
   & Scrappie (R10.4.1) & \ding{51} & \textcolor{teal}{+0.048} & \textcolor{teal}{+0.004} & \textcolor{teal}{+0.064} & \textcolor{teal}{-6,841.54} \\
   & Scrappie (R10.4.1) & \ding{55} & \textcolor{teal}{+0.105} & \textcolor{teal}{+0.021} & \textcolor{teal}{+0.128} & \textcolor{teal}{-5,391.66} \\
\bottomrule
\end{tabular}
\end{table}

\vspace{1em}

\subsection{Ablation Results for Error Correction}

In Supplementary Table~\ref{tab:correction-delta} we report the effect of \mech's error correction mechanism on mapping quality and performance for each dataset and configuration. That is, for each configuration of segmentation method and HPC we report the difference between using and not using error correction. Improvements in a metric are displayed in teal, regressions in red.

\vspace{1em}

\begin{table}[htbp]
\centering
\caption{Effect of error correction: difference (corrected $-$ uncorrected) for each configuration.}
\label{tab:correction-delta}
\small
\begin{tabular}{ll c rrrr}
\toprule
Dataset & Segmenter & HPC & $\Delta$F1 & $\Delta$Precision & $\Delta$Recall & $\Delta$Time to Map (ms) \\
\midrule
  \textit{E.~coli} & Campolina & \ding{55} & \textcolor{teal}{+0.011} & \textcolor{teal}{+0.004} & \textcolor{teal}{+0.014} & \textcolor{red}{+0.29} \\
   & Campolina & \ding{51} & \textcolor{teal}{+0.009} & \textcolor{teal}{+0.000} & \textcolor{teal}{+0.014} & \textcolor{red}{+0.78} \\
   & Scrappie (R9.4) & \ding{55} & \textcolor{teal}{+0.550} & \textcolor{teal}{+0.236} & \textcolor{teal}{+0.477} & \textcolor{teal}{-3.80} \\
   & Scrappie (R9.4) & \ding{51} & \textcolor{teal}{+0.126} & \textcolor{teal}{+0.016} & \textcolor{teal}{+0.162} & \textcolor{teal}{-0.12} \\
   & Scrappie (R10.4.1) & \ding{55} & \textcolor{teal}{+0.041} & \textcolor{teal}{+0.017} & \textcolor{teal}{+0.042} & \textcolor{red}{+0.29} \\
   & Scrappie (R10.4.1) & \ding{51} & \textcolor{teal}{+0.020} & \textcolor{teal}{+0.001} & \textcolor{teal}{+0.028} & \textcolor{red}{+0.52} \\
\midrule
  \textit{D.~melanogaster} & Campolina & \ding{55} & \textcolor{teal}{+0.004} & \textcolor{teal}{+0.016} & \textcolor{red}{-0.004} & \textcolor{teal}{-48.60} \\
   & Campolina & \ding{51} & \textcolor{teal}{+0.010} & +0.000 & \textcolor{teal}{+0.017} & \textcolor{red}{+41.65} \\
   & Scrappie (R9.4) & \ding{55} & \textcolor{teal}{+0.519} & \textcolor{teal}{+0.387} & \textcolor{teal}{+0.515} & \textcolor{teal}{-1,722.36} \\
   & Scrappie (R9.4) & \ding{51} & \textcolor{teal}{+0.085} & \textcolor{teal}{+0.003} & \textcolor{teal}{+0.129} & \textcolor{red}{+152.44} \\
   & Scrappie (R10.4.1) & \ding{55} & \textcolor{teal}{+0.014} & \textcolor{teal}{+0.000} & \textcolor{teal}{+0.019} & \textcolor{red}{+35.52} \\
   & Scrappie (R10.4.1) & \ding{51} & \textcolor{teal}{+0.012} & \textcolor{red}{-0.003} & \textcolor{teal}{+0.019} & \textcolor{red}{+121.94} \\
\midrule
  \textit{H.~sapiens} & Campolina & \ding{55} & \textcolor{teal}{+0.003} & \textcolor{teal}{+0.001} & \textcolor{teal}{+0.004} & \textcolor{red}{+782.84} \\
   & Campolina & \ding{51} & \textcolor{teal}{+0.007} & +0.000 & \textcolor{teal}{+0.012} & \textcolor{red}{+396.91} \\
   & Scrappie (R9.4) & \ding{55} & \textcolor{teal}{+0.773} & \textcolor{teal}{+0.982} & \textcolor{teal}{+0.635} & \textcolor{teal}{-9,697.16} \\
   & Scrappie (R9.4) & \ding{51} & \textcolor{teal}{+0.762} & \textcolor{teal}{+0.147} & \textcolor{teal}{+0.659} & \textcolor{red}{+497.62} \\
   & Scrappie (R10.4.1) & \ding{55} & \textcolor{teal}{+0.073} & \textcolor{teal}{+0.017} & \textcolor{teal}{+0.086} & \textcolor{red}{+2,093.18} \\
   & Scrappie (R10.4.1) & \ding{51} & \textcolor{teal}{+0.015} & \textcolor{teal}{+0.001} & \textcolor{teal}{+0.022} & \textcolor{red}{+643.30} \\
\bottomrule
\end{tabular}
\end{table}

\clearpage

\subsection{Parameter Search}

Supplementary Tables~\ref{tab:no_hpc_parameter_search} and~\ref{tab:hpc_parameter_search} report the parameter search results for each segmentation method with HPC disabled and enabled, respectively. For each number of HMM states, we report the parameter search runtime, the selected stay and skip probabilities, and the resulting F1 score. Across the tables, the best F1 score for each configuration is shown in bold, while the second-best result is underlined.

\begin{table}[h]
\centering
\caption{Parameter search when HPC is disabled.}
\label{tab:no_hpc_parameter_search}
\begin{tabular}{lrrrrrrrrr}
\toprule
Number of States & Baseline & 8 & 16 & 32 & 64 & 96 & 128 & 192 & 256 \\
\midrule
\multicolumn{10}{c}{\textbf{Scrappie R9}} \\
\midrule
Parameter Search Runtime & - & 826 & 691 & 652 & 658 & 644 & 713 & 705 & 839 \\
P(stay) & - & 0.250 & 0.270 & 0.280 & 0.275 & 0.340 & 0.300 & 0.280 & 0.315 \\
P(skip) & - & 0.010 & 0.000 & 0.000 & 0.000 & 0.000 & 0.020 & 0.000 & 0.005 \\
F1 & 0.104 & 0.596 & 0.730 & 0.754 & 0.784 & 0.787 & \underline{0.792} & 0.788 & \textbf{0.794} \\
\midrule
\multicolumn{10}{c}{\textbf{Scrappie R10}} \\
\midrule
Parameter Search Runtime & - & 535 & 470 & 488 & 481 & 462 & 493 & 573 & 545 \\
P(stay) & - & 0.005 & 0.025 & 0.035 & 0.060 & 0.050 & 0.050 & 0.040 & 0.050 \\
P(skip) & - & 0.025 & 0.010 & 0.005 & 0.010 & 0.005 & 0.020 & 0.035 & 0.040 \\
F1 & 0.699 & 0.711 & 0.728 & 0.737 & 0.741 & \textbf{0.744} & \underline{0.743} & 0.742 & 0.741 \\
\midrule
\multicolumn{10}{c}{\textbf{Campolina}} \\
\midrule
Parameter Search Runtime & - & 508 & 511 & 573 & 545 & 544 & 534 & 519 & 580 \\
P(stay) & - & 0.015 & 0.005 & 0.025 & 0.020 & 0.030 & 0.035 & 0.050 & 0.025 \\
P(skip) & - & 0.000 & 0.000 & 0.045 & 0.005 & 0.025 & 0.010 & 0.050 & 0.030 \\
F1 & 0.863 & 0.868 & 0.876 & 0.873 & 0.875 & 0.874 & \underline{0.876} & 0.875 & \textbf{0.876} \\
\bottomrule
\end{tabular}
\end{table}

\vspace{1em}

\begin{table}[h]
\centering
\caption{Parameter search when HPC is enabled.}
\label{tab:hpc_parameter_search}
\begin{tabular}{lrrrrrrrrr}
\toprule
Number of States & Baseline & 8 & 16 & 32 & 64 & 96 & 128 & 192 & 256 \\
\midrule
\multicolumn{10}{c}{\textbf{Scrappie R9}} \\
\midrule
Parameter Search Runtime & - & 854 & 747 & 897 & 783 & 797 & 893 & 987 & 1024 \\
P(stay) & - & 0.200 & 0.200 & 0.190 & 0.215 & 0.290 & 0.325 & 0.290 & 0.325 \\
P(skip) & - & 0.045 & 0.000 & 0.105 & 0.000 & 0.000 & 0.025 & 0.030 & 0.005 \\
F1 & 0.727 & 0.832 & \textbf{0.866} & 0.859 & 0.863 & 0.860 & 0.863 & 0.862 & \underline{0.865} \\
\midrule
\multicolumn{10}{c}{\textbf{Scrappie R10}} \\
\midrule
Parameter Search Runtime & - & 779 & 581 & 565 & 530 & 548 & 519 & 541 & 582 \\
P(stay) & - & 0.010 & 0.010 & 0.000 & 0.000 & 0.000 & 0.000 & 0.000 & 0.000 \\
P(skip) & - & 0.010 & 0.000 & 0.010 & 0.000 & 0.000 & 0.000 & 0.000 & 0.000 \\
F1 & 0.830 & \underline{0.849} & \textbf{0.852} & 0.844 & 0.847 & 0.842 & 0.841 & 0.841 & 0.839 \\
\midrule
\multicolumn{10}{c}{\textbf{Campolina}} \\
\midrule
Parameter Search Runtime & - & 752 & 657 & 666 & 705 & 660 & 583 & 702 & 645 \\
P(stay) & - & 0.005 & 0.000 & 0.000 & 0.005 & 0.000 & 0.000 & 0.005 & 0.000 \\
P(skip) & - & 0.080 & 0.015 & 0.080 & 0.105 & 0.055 & 0.000 & 0.055 & 0.000 \\
F1 & 0.910 & 0.912 & \textbf{0.919} & \underline{0.916} & 0.915 & 0.915 & 0.915 & 0.914 & 0.915 \\
\bottomrule
\end{tabular}
\end{table}

\clearpage

\subsection{Ablation Results for Stay Removal and Noise Reduction}
Supplementary Table~\ref{tab:ablation_parameter_search} presents an ablation analysis testing the effects of \mech with only stay removal, and \mech with only noise removal. A parameter sweep across HMM sizes and segmentation algorithms was performed for both configuration. All parameter searches were guided by RawHash2 with HPC off.
\vspace{1em}
\begin{table}[h]
\centering
\caption{Ablation parameter search.}
\label{tab:ablation_parameter_search}
\begin{tabular}{lrrrrrrrrr}
\toprule
Number of States & Baseline & 8 & 16 & 32 & 64 & 96 & 128 & 192 & 256 \\
\midrule
\multicolumn{10}{c}{\textbf{Scrappie R9  (Only Stay Removal)}} \\
\midrule
Parameter Search Runtime & - & 843 & 724 & 762 & 759 & 668 & 815 & 833 & 920 \\
P(stay) & - & 0.220 & 0.230 & 0.255 & 0.250 & 0.300 & 0.340 & 0.290 & 0.320 \\
P(skip) & - & 0.005 & 0.050 & 0.005 & 0.005 & 0.025 & 0.010 & 0.005 & 0.030 \\
F1 & 0.104 & 0.604 & 0.717 & 0.747 & 0.775 & 0.780 & \underline{0.785} & 0.783 & \textbf{0.787} \\
\midrule
\multicolumn{10}{c}{\textbf{Scrappie R10  (Only Stay Removal)}} \\
\midrule
Parameter Search Runtime & - & 503 & 532 & 660 & 540 & 562 & 534 & 506 & 566 \\
P(stay) & - & 0.025 & 0.055 & 0.035 & 0.055 & 0.040 & 0.050 & 0.050 & 0.045 \\
P(skip) & - & 0.070 & 0.070 & 0.020 & 0.025 & 0.005 & 0.035 & 0.035 & 0.025 \\
F1 & 0.699 & 0.701 & 0.710 & 0.724 & 0.732 & \textbf{0.738} & 0.735 & \underline{0.736} & 0.734 \\
\midrule
\multicolumn{10}{c}{\textbf{Campolina  (Only Stay Removal)}} \\
\midrule
Parameter Search Runtime & - & 443 & 553 & 519 & 534 & 571 & 505 & 518 & 625 \\
P(stay) & - & 0.000 & 0.015 & 0.025 & 0.025 & 0.020 & 0.025 & 0.025 & 0.025 \\
P(skip) & - & 0.000 & 0.010 & 0.050 & 0.210 & 0.180 & 0.100 & 0.025 & 0.185 \\
F1 & 0.863 & 0.863 & 0.864 & 0.866 & 0.868 & 0.868 & 0.869 & \underline{0.870} & \textbf{0.870} \\
\midrule
\multicolumn{10}{c}{\textbf{Scrappie R9  (Only Noise Removal)}} \\
\midrule
Parameter Search Runtime & - & 980 & 865 & 898 & 912 & 917 & 914 & 888 & 907 \\
P(stay) & - & 0.320 & 0.300 & 0.200 & 0.100 & 0.120 & 0.105 & 0.035 & 0.000 \\
P(skip) & - & 0.055 & 0.300 & 0.030 & 0.290 & 0.280 & 0.090 & 0.000 & 0.120 \\
F1 & 0.104 & 0.101 & \textbf{0.112} & 0.108 & 0.110 & \underline{0.110} & 0.108 & 0.109 & 0.107 \\
\midrule
\multicolumn{10}{c}{\textbf{Scrappie R10  (Only Noise Removal)}} \\
\midrule
Parameter Search Runtime & - & 579 & 525 & 533 & 479 & 555 & 498 & 527 & 551 \\
P(stay) & - & 0.310 & 0.195 & 0.275 & 0.100 & 0.060 & 0.335 & 0.195 & 0.320 \\
P(skip) & - & 0.025 & 0.000 & 0.000 & 0.000 & 0.015 & 0.000 & 0.000 & 0.000 \\
F1 & 0.699 & 0.716 & \textbf{0.723} & 0.717 & \underline{0.718} & 0.713 & 0.713 & 0.712 & 0.714 \\
\midrule
\multicolumn{10}{c}{\textbf{Campolina  (Only Noise Removal)}} \\
\midrule
Parameter Search Runtime & - & 618 & 587 & 557 & 499 & 537 & 573 & 563 & 577 \\
P(stay) & - & 0.330 & 0.030 & 0.000 & 0.000 & 0.060 & 0.270 & 0.005 & 0.000 \\
P(skip) & - & 0.005 & 0.000 & 0.080 & 0.000 & 0.000 & 0.000 & 0.000 & 0.050 \\
F1 & 0.863 & 0.870 & \textbf{0.877} & 0.872 & 0.872 & 0.870 & \underline{0.873} & 0.871 & 0.869 \\
\bottomrule
\end{tabular}
\end{table}

\let\noopsort\undefined
\let\printfirst\undefined
\let\singleletter\undefined
\let\switchargs\undefined

\end{document}